\documentclass[entropy,article,accept,pdftex,moreauthors]{Definitions/mdpi} 

\usepackage{braket}
\definecolor{darkred}{rgb}{0.90,0.2,0.2}
\definecolor{darkgreen}{rgb}{0,0.60,.2}
\definecolor{darkblue}{rgb}{0.1,0.3,1}
\definecolor{grey}{cmyk}{0,0,0,0.25}
\definecolor{orange}{cmyk}{0,0.6,0.8,0}

\firstpage{1} 
\pubvolume{1}
\issuenum{1}
\articlenumber{0}
\pubyear{2024}
\copyrightyear{2024}
\datepublished{ } 
\hreflink{https://doi.org/} 

\Title{Survival Probability, Particle Imbalance, and Their Relationship in Quadratic Models}

\TitleCitation{Survival Probability, Particle Imbalance, and Their Relationship in Quadratic Models}

\Author{{Miroslav} 
 Hopjan $^{1,}${*}
\orcidA{} and Lev Vidmar $^{1,2}$\orcidB{}}

\AuthorNames{Miroslav Hopjan and Lev Vidmar}

\AuthorCitation{{Hopjan}
, M.; Vidmar, L.}

\address{%
$^{1}$ \quad Department of Theoretical Physics, J. Stefan Institute, SI-1000 Ljubljana, Slovenia; {lev.vidmar@ijs.si} 
 \\
$^{2}$ \quad Department of Physics, Faculty of Mathematics and Physics, University of Ljubljana, \mbox{SI-1000 Ljubljana, Slovenia}}

\corres{Correspondence: miroslav.hopjan@ijs.si
}

\abstract{
We argue that the dynamics of particle imbalance in quadratic fermionic models is, for the majority of initial many-body product states  {in the} 
 site occupation basis, 
virtually indistinguishable from the dynamics of survival probabilities of single-particle states.
We then generalize our statement to a similar relationship between the non-equal time and space density correlation functions in many-body states, and the transition probabilities of single-particle states at nonzero distances.
Finally, we study the equal-time connected density--density correlation functions in many-body states, which exhibit certain qualitative analogies with the survival and transition probabilities of single-particle states. 
Our results are numerically tested for two paradigmatic models of single-particle localization: the 3D Anderson model and the 1D Aubry--André model. 
This work gives an affirmative answer to the question of whether it is possible to measure features of single-particle survival and transition probabilities by the dynamics of observables in many-body states.
}


\begin{document}

\section{Introduction}
\label{sec:intro}

The survival probability of an initial state $|j\rangle$ is defined as the square of its overlap with the time-evolving quantum state $|j(t)\rangle = e^{-i\hat Ht}|j\rangle$ under the Hamiltonian $\hat H$,
\begin{equation}\label{eq:first}
    P_{jj}^{\,H}(t) = |\langle j| e^{-i\hat Ht}|j\rangle|^2 \;,
\end{equation}
{where} 
 we set $\hbar = 1$.
It represents a useful probe to study the dynamical properties of $\hat H$ and, hence, it is of broad interest in the theory of quantum chaos and ergodicity breaking phenomena~\cite{Ketzmerick_92, Huckestein94, Schofield_95, Schofield_96, Brandes96, Ketzmerick_97, Ohtsuki97, Gruebele_98, Ng_06, torresherrera_santos_14, torresherrera_santos_15, Leitner_15, Santos2017, torresherrera_garciagarcia_18, Bera2018, Prelovsek18, Schiulaz_19, Karmakar_19, Lezama_21, hopjan2023, hopjan2023scaleinvariant, das2024proposal}.
For example, it was shown that the averaged survival probability,
\begin{align}
    \label{eq:def_Pd}
    P_{}^{\,H}(t) = \braket{\,P_{jj}^{\,H}(t)\,}_j \;,
\end{align}
represents a particularly useful tool for the detection of eigenstate transitions~\cite{hopjan2023, hopjan2023scaleinvariant}.
The average $\braket{\dots}_j$ in Equation~(\ref{eq:def_Pd}) is carried out over all possible initial states $|j\rangle$ that can be thought of as the eigenstates of the Hamiltonian $\hat{H}_0$ before the quench.
The eigenstate transitions may correspond to single-particle localization transitions in eigenstates of $\hat{H}$ when the initial states $|j\rangle$ are single-particle states, or ergodicity breaking phase transitions when the initial states $|j\rangle$ are many-body states.

Here, we study quadratic fermionic models and search for quantitative similarities between the dynamics of single-particle quantities, such as survival probability and the dynamics of observables in many-body states.
Recently, we reported the observation that the dynamics of site occupations and particle imbalance (to be defined below) exhibit critical behavior at the localization transition point~\cite{jiricek2023critical}, similar to the critical behavior of survival probability~\cite{hopjan2023, hopjan2023scaleinvariant}.
This observation corresponds to the dynamics of the initial many-body states that are charge density-wave (CDW) product states at half filling, i.e., the states that are routinely studied in experiments~\cite{Schreiber15, Choi16, Luschen17, Bordia17, Kohlert19, Rubio-Abadal19,
Guo20}.
{The} 
 property of the CDW product states is that the neighboring sites of an occupied lattice site are always empty, at least along a selected direction.
The particle imbalance is defined as the normalized sum of weighted site occupations $n^{\,H}_j(t)$ over the entire lattice with $V$ sites,
\begin{equation} \label{eq:def_I_intro}
  I_{}^{\,H}(t)= \frac{2}{V}\sum_{j=1}^V (-1)^{n_j(0)-1} n^{\,H}_j(t)\;,\;\;\;{\rm with}\;\;\;
  n^{\,H}_j(t) = \langle \Psi_0| e^{i\hat H t} \hat n_j e^{-i\hat H t} |\Psi_0\rangle\;,
\end{equation}
where $\hat n_j = \hat c_j^\dagger \hat c_j$ is the site-occupation (density) operator, $|\Psi_0 \rangle$ is the initial many-body state, $n_j(0)$ is the site occupation at time $t=0$, and the prefactor $2/V$ assures the unit normalization at $t=0$.
The imbalance can also be expressed as the mean of the site-occupation (density) autocorrelation functions,
\begin{equation} \label{eq:autocorrel}
I_{}^{\,H}(t)= \frac{4}{V}\sum_{j=1}^V \Big[n^{\,H}_j(t)-1/2\Big] \Big[n_j(0)-1/2\Big] \;.
\end{equation}
The property that one may detect the eigenstate transitions via the dynamics of particle imbalance~\cite{jiricek2023critical} shares similarities with the detection of eigenstate transitions via the survival probability~\cite{hopjan2023, hopjan2023scaleinvariant}.
Intriguingly, even the exponent $\beta_I$ of the power-law decay of imbalance at criticality from the initial CDW product states is quantitatively close to the decay exponent $\beta$ of the survival probability in the 3D Anderson model, though it is dissimilar in the 1D Aubry--André model.
Similarities between the density-wave imbalance and the survival probability were also studied in delocalized disordered systems~\cite{Popperl22}.
These results call for a more detailed understanding of the connection of the dynamics of density correlations in many-body states with the dynamics of single-particle quantities such as survival probabilities.

In this work, we go beyond Ref.~\cite{jiricek2023critical} and focus on the quantitative comparison between the dynamics in many-body states and single-particle states, ranging from short to long times, considering initial many-body product states on a site occupation basis without the CDW order, as well as tuning the disorder strength to the localization transition point and away from it.
Our study is carried out while having in mind the 3D Anderson model and 1D Aubry--André model (to be introduced in Section \ref{sec:models}); however, we expect our results to generally apply to quadratic models with quenched disorder and localization transitions.
The main result of this work is that for the overwhelming majority of initial many-body product states on a site occupation basis, dubbed typical product states, the dynamics of particle imbalance are virtually indistinguishable from the dynamics of single-particle survival probabilities,
\begin{equation} \label{eq:main}
  I_{}^{\,H}(t)\approx P_{}^{\,H}(t)\;.
\end{equation}
The derivation of Equation \eqref{eq:main} and its numerical tests are carried out in Section \ref{sec:imbalance}.
In Section~\ref{sec:imbalance_correlations}, we generalize this result to a similar relationship between the non-equal time and space density correlation functions and the single-particle transition probabilities between lattice sites at non-zero distances. 
In Section \ref{sec:correlations_equal}, we discuss qualitative similarities between the dynamics of equal-time connected density--density correlation functions and the dynamics of single-particle survival and transition probabilities.
We conclude in Section~\ref{sec:discussion}.


\section{Models} \label{sec:models}

We consider two paradigmatic quadratic models of fermions without spin structure that exhibit single-particle localization  transitions, the Anderson model~\cite{anderson_58, Abrahams79,evers_mirlin_08} and the Aubry--André model~\cite{Aubry80, Suslov82}.
The models are given by the Hamiltonian

\begin{equation}
\label{eq:hamnonint}
\hat H= -{J}\sum_{\langle ij\rangle}^{} (\hat{c}_{i}^{\dagger}\hat{c}_{j}^{}+ \hat c_j^\dagger \hat c_i) + \sum_{i=1}^{V}\epsilon_{i}\hat{n}_{i}^{}\;,
\end{equation}
where $\langle ij\rangle$ denotes nearest neighbors, $J$ stands for the strength of hopping matrix element linking the nearest neighbors, $\hat{c}_{i}^{\dagger}$ ($\hat{c}_{i}^{}$) are the fermionic creation (annihilation) operators at site $i$,  $\hat{n}_{i}^{}=\hat{c}_{i}^{\dagger}\hat{c}_{i}^{}$ is the site occupation (density) operator, and $\epsilon_{i}$ represents the on-site energy.
The number of lattice sites is denoted as $V$, which also equals the single-particle Hilbert space dimension $D$, i.e., $D=V$.

As the first example, we inspect the Anderson model on a three-dimensional (3D) cubic lattice of volume $V=L^3$, where $L$ is the linear size, with periodic boundary conditions. 
The on-site energies $\epsilon_i$ are independently and identically distributed, their values are taken out from a box distribution $\epsilon_i \in [-W/2,W/2]${.} 
{Properties of the 3D Anderson model have been discussed in several reviews~\cite{kramer_mackinnon_93,brandes2003anderson,evers_mirlin_08,Lagendijk2009, suntajs_prosen_21}; below, we summarize some of them that are relevant for this study.}
The position of the localization critical point and its properties were discussed from different perspectives~\cite{kramer_mackinnon_93, MacKinnon81, MacKinnon83, suntajs_prosen_21, Tarquini2017}, with the broad acceptance that the system is insulating---i.e., with all single-particle eigenstates localized---for $W >W_{c} \approx 16.5\,J$~\cite{Ohtsuki18}. For $W<W_{c}$, the transport is dominated by diffusive eigenstates~\cite{Ohtsuki97,Zhao20,Herbrych21}. At the critical point, the diffusive eigenstates vanish and the multifractal eigenstates~\cite{evers_mirlin_08, Rodriguez09, Rodriguez10} govern the transport that then becomes subdiffusive~\cite{Ohtsuki97}.
Moreover, below the critical disorder $W <W_{c}$, mobility edges in the spectrum separate the localized eigenstates from the delocalized ones. The mobility edges shift towards the band edges when disorder is decreased~\cite{Schubert_05}.
Whereas the localization transition is typically studied within the framework of single-particle properties, its manifestations can also be detected in many-body states~\cite{Li16, hopjan_orso_21, Zhao20, bhakuni2023dynamic,jiricek2023critical}. 

The second model that we inspect is the Aubry--André model  on a one-dimensional (1D) lattice of size $L$ with closed boundary conditions. 
In this model, the quasiperiodic on-site potential $\epsilon_{i}=\lambda \cos(2\pi q i+\phi)$ is imposed on the lattice, $\lambda$ represents the amplitude of the potential, and $\phi$ is a global phase. The  periodicity of the potential is incommensurate with the periodicity of the lattice by the standard choice of the golden ratio value $q=\frac{\sqrt{5}-1}{2}$. 
{Its properties were discussed in Ref.~\cite{Dominguez-Castro_2019}, and below we limit the discussion to those that are relevant for this study.
}
At $\lambda_c = 2J$, the 1D Aubry--André model displays an abrupt transition from delocalized to localized phase~\cite{Aubry80, Suslov82, Kohmoto83, Chao86, Kohmoto87, Siebesma1987, Hiramoto89, Hiramoto92, Macia2014, Li16, wu2021}. The model shows self-dual property. On the one hand, at $\lambda>\lambda_c$, all states are localized in real space with delocalization exhibited in momentum space; on the other hand, at $\lambda<\lambda_c$, all states are delocalized in real space with localization exhibited in momentum space. At $\lambda=\lambda_c$, i.e., the critical point, both the eigenspectrum and eigenstates are (multi)fractal, and the model exhibits diffusion$~$\cite{Geisel_91} or atypical scaling $\propto L^2$  of the typical Heisenberg time$~$\cite{hopjan2023}. The latter can be understood as a remnant two-dimensionality of the 1D Aubry--André model. Indeed, the model is closely associated to the Harper--Hofstadter model, which describes an electron moving in an isotropic 2D lattice subjected to magnetic field$~$\cite{Harper_1955}. The transition was experimentally realized using photonic lattices~\cite{Lahini:PRL2009} and cold atoms~\cite{Roati08,Luschen18}. 
As in the Anderson model, the localization transition in the 1D Aubry--André model is typically studied within the framework of single-particle properties; however, its manifestations can also be detected in many-body states~\cite{Li16, DeTomasi21, hopjan_orso_21, Roy2021a, Roy2021b, aditya2023familyvicsek,jiricek2023critical}.


\section{Survival Probability and Particle Imbalance}
\label{sec:imbalance}

We consider the following quench protocol.
The initial Hamiltonian is $\hat{H}_0=\sum_{i=1}^{V}\epsilon_{i}\hat{n}_{i}^{}\;,$ which can be thought of as the limit of infinite-strength disorder, and the final Hamiltonian $\hat{H}$ is given by Equation \eqref{eq:hamnonint}.
The initial many-body states $|\Psi_0\rangle$ are eigenstates of $\hat{H}_0$ and can be written as product states on a site occupation basis,
\begin{equation} \label{eq:def_initial}
    |\Psi_0\rangle = \prod_{j_l  \in \Psi_0} \hat c_{j_l}^\dagger |\emptyset\rangle \;,
\end{equation}
where the product runs over the sites $j_l$ that are occupied in the initial states (we consider half filling, i.e., the number of particles is $N=V/2$).
The time evolution of the particle occupation $n_j^H$ at site $j$ can be expressed as~\cite{jiricek2023critical}
\begin{equation} \label{eq:def_ni_1}
n^{\,H}_j(t) = \sum_{\substack{j_l \in \Psi_0}}^{}  P_{jj_l}^{\,H}(t)\;,
\end{equation}
where we introduce the time evolution of single-particle transition probabilities
\begin{equation} \label{eq:def_trans}
P_{jj_l}^{\,H}(t)=|\langle j| e^{-i\hat Ht}|j_l\rangle|^2\;,
\end{equation}
from the initially occupied single-particle state $|j_l\rangle$ to the state $|j\rangle$.
We stress that the simple relation in Equation \eqref{eq:def_ni_1} between the time evolution of observable in many-body states and the transition probabilities of single-particle states applies only for the observables that share the common eigenbasis with the Hamiltonian $\hat{H}_0$ before the quench---see the detailed derivation in Ref.~\cite{jiricek2023critical}.

Equation$~$\eqref{eq:def_ni_1} allows one to explicitly connect the dynamics of site occupations in many-body states with the single-particle survival probability.
For the initially occupied sites, $\substack{j \in \Psi_0}$, the contribution of the single-particle survival probability can be taken out of the sum in Equation \eqref{eq:def_ni_1},
\begin{equation} \label{eq:def_ni_2}
n^{\,H}_{j\in \Psi_0}(t) = P_{jj}^{\,H}(t) + \sum_{\substack{j_l \in \Psi_0}}^{j_l\neq j}  P_{jj_l}^{\,H}(t)\;.
\end{equation}
The time evolution of the particle imbalance $I_{}^{\,H}(t)$ in Equation \eqref{eq:def_I_intro} can then be expressed only via the site occupations of the initially occupied sites,
{\small
\begin{align} \label{eq:def_imbalance2}
    I_{}^{\,H}(t) = \frac{2}{V} \sum_{j\in\Psi_0} n_j^H(t) - \frac{2}{V} \sum_{j\notin\Psi_0} n_j^H(t) 
    = \big\langle n^{\,H}_j(t)  \big\rangle_{j\in\Psi_0} - \big\langle n^{\,H}_j(t)  \big\rangle_{j\notin\Psi_0} 
    = 2\big\langle n^{\,H}_j(t)  \big\rangle_{j\in\Psi_0} -1\;,
\end{align}
}
where we define the average over initially occupied sites $\big\langle ...  \big\rangle_{j\in\Psi_0} = \frac{2}{V} \sum_{j\in\Psi_0} ...$ (with $j\in\Psi_0 \to j\notin\Psi_0$ for the average over initially unoccupied sites) and we used the particle sum rule $\big\langle n^{\,H}_j(t)\big\rangle_{j\in\Psi_0} + \big\langle n^{\,H}_j(t)  \big\rangle_{j\notin\Psi_0} =1$.

Equation$~$\eqref{eq:def_imbalance2} provides the basis for our derivation of Equation \eqref{eq:main}. To this end, we substitute Equation \eqref{eq:def_ni_2} into Equation \eqref{eq:def_imbalance2} and separate the contribution from the survival probabilities as
\begin{align} \label{eq:imbal_surv}
    I_{}^{\,H}(t) = \big\langle\, P^{\,H}_{jj}(t)\,\big\rangle_{j\in\Psi_0}\, + \big\langle\, P^{\,H}_{jj}(t)\,+ 2\sum_{\substack{j_l \in \Psi_0}}^{j_l\neq j}  P^{\,H}_{jj_l}(t) \,\big\rangle_{j\in\Psi_0}\,-1\;.
\end{align}
This equation suggests that the particle imbalance and survival probability become identical if the second and the third term on the r.h.s.~of Equation \eqref{eq:imbal_surv} cancel, i.e., if
\begin{align} \label{eq:equivalence}
\big\langle\, P^{\,H}_{jj}(t)\,+ 2\sum_{\substack{j_l \in \Psi_0}}^{j_l\neq j}  P^{\,H}_{jj_l}(t) \,\big\rangle_{j\in\Psi_0}\,= 1\,.
\end{align}
However, this is in general not the case and, hence, the formal equivalence between the particle imbalance and survival probabilities is not expected to hold.

At this point, one can ask whether there exist initial states for which the particle imbalance and survival probabilities still become approximately identical.
We argue that this is indeed the case for the overwhelming majority of initial many-body states under consideration.
We refer to them as typical initial product states.
At half filling, the sites $j$ of a typical initial product state have their neighbors occupied or not occupied with equal probability.
This gives rise to the self-averaging property of the wavefunction, since the sum on the l.h.s.~of Equation \eqref{eq:equivalence} can be expressed as

\begin{align} \label{eq:equivalence2}
\big\langle\, P^{\,H}_{jj}(t)+ 2\sum_{\substack{j_l \in \Psi_0}}^{j_l\neq j}  P^{\,H}_{jj_l}(t) \big\rangle_{j\in\Psi_0}\,
\approx 
\big\langle P^{\,H}_{jj}(t)+ \frac{2}{2}\sum_{\substack{i=1\\ i\neq j}}^{V} P^{\,H}_{ji}(t) \big\rangle_{j\in\Psi_0}
= \big\langle \sum_{i}^{}  P^{\,H}_{ji}(t) \big\rangle_{j\in\Psi_0} = 1 \;,
\end{align}
where in the last step we have used the conservation of all transition probabilities, including the survival probability $\sum_{i}^{}  P^{\,H}_{ji}(t)=1$. 
Thus, we arrive at the approximate equivalence,\vspace{-6pt}
\begin{align}  \label{eq:equivalence_half}
    I_{}^{\,H}(t) \approx \big\langle\, P^{\,H}_{jj}(t)\,\big\rangle_{j\in\Psi_0}\,,
\end{align}
which is expected to be valid for a typical initial state.
Note that in Equation (\ref{eq:equivalence_half}) the average $\big\langle ... \big\rangle_{j\in\Psi_0}$ is carried out over the initially occupied sites, i.e., over $1/2$ of all lattice sites, while the average $\big\langle ... \big\rangle_{j}$ in Equation (\ref{eq:def_Pd}) is carried out over all lattice sites.
The difference between these two definitions of averaging is insignificant, which was numerically confirmed in~\cite{jiricek2023critical}.
This allows us to finally derive Equation \eqref{eq:main},\vspace{-6pt}
\begin{align} \label{eq:equivalence4}
    I_{}^{\,H}(t) \approx \big\langle\, P^{\,H}_{jj}(t)\,\big\rangle_{j\in\Psi_0}\,\approx \big\langle\, P^{\,H}_{jj}(t)\,\big\rangle_{j}\,=P_{}^{\,H}(t)\;,
\end{align}
which is the main result of this work.
We stress that this result is derived considering a single Hamiltonian realization.
Still, in the actual numerical calculations in finite systems we also carry out the average over different realizations of Hamiltonians after the quench,
$I_{}^{\,}(t) = \big\langle\, I_{}^{H\,}(t)\,\big\rangle_{H}\,$ and $P_{}^{\,}(t)=\big\langle\, P_{}^{H\,}(t)\,\big\rangle_{H}\,$, such that Equation (\ref{eq:equivalence4}) is rewritten to\vspace{-6pt}
\begin{align} \label{eq:equivalence5}
    I_{}^{\,}(t) \approx P_{}^{\,}(t)\;.
\end{align}
The above result suggests that the time evolution of the particle imbalance from a typical initial product state is approximately identical to the time evolution of the survival probability.
The overwhelming majority of eigenstates of $\hat H_0$ belong to this category.
However, the experimentally relevant initial product states, which form a CDW pattern, do not belong to this category and, hence, they can be considered as atypical states.
The evidence for the initial CDW states violating Equation (\ref{eq:equivalence5}) was shown in Ref.~\cite{jiricek2023critical} for the 1D Aubry--André model at the critical point, at which the exponent of the power-law decay of $I_{}(t)$ did not match the exponent of $P_{}(t)$.

We next test our results numerically.
We generate initial states by randomly selecting one of the many-body eigenstates of the initial Hamiltonian $\hat{H}_0$. 
The overwhelming majority of many-body eigenstates are expected to be typical in the sense defined here.
To decrease the effect of rare atypical states, especially for small system sizes, we further average over 50 different Hamiltonian realizations (i.e., over the on-site energies $\epsilon_i$ in the 3D Anderson model and over the global phase $\phi$ in the 1D Aubry--André model).

\begin{figure}[H]
\includegraphics[width=6.3 cm]{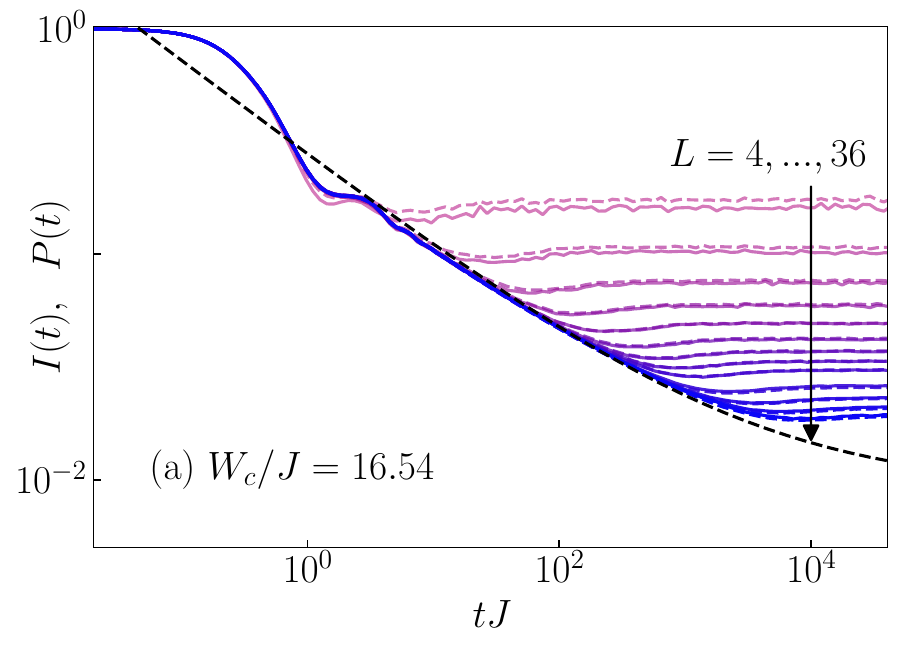}
\includegraphics[width=6.3 cm]{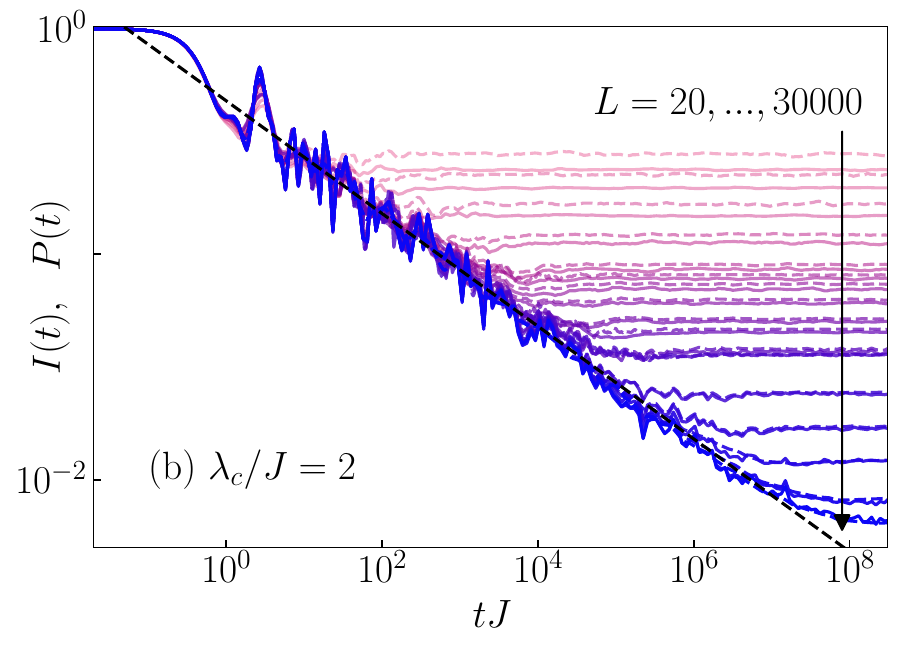}
\vspace{-0.2cm}
\caption{
\textls[-15]{{Dynamics} 
 of the survival probability $P(t)$ (dashed lines) and particle imbalance $I(t)$ (solid lines), see Equation (\ref{eq:equivalence5}), from the typical initial product states.
Results are shown for (\textbf{a}) the 3D Anderson model at the critical point \mbox{$W_c/J = 16.5$} and for system sizes $L = 4,6,8,10,12,14,16,18,20,24,28,32,36$, and \mbox{(\textbf{b}) the 1D} Aubry--André model at the critical point $\lambda_c/J=2$ and for system {sizes} 
 $L =$ 20, 30, 50, 100, 200, 250, 300, 400, 500, 750, 1000, 1250, 2500, 5000, 10,000, 20,000, 30,000. The black dashed line is a fit to the function $a (tJ)^{-\beta} + P_\infty$, where $P_\infty$ is the infinite-time value of $P(t)$ in the thermodynamic limit, which is non-zero (zero) in the case of the 3D Anderson model (1D Aubry--André model) due to the existence (absence) of the mobility edge, see also~\cite{hopjan2023, hopjan2023scaleinvariant}.  }
}
\label{fig1}
\end{figure}

In Figure \ref{fig1}a,b, we compare ${I}(t)$ with $P^{}(t)$ for the 3D Anderson model and the 1D Aubry--André model, respectively, at the corresponding eigenstate transition points.
We observe that the time evolution of the survival probability $P^{}(t)$ is indeed very close to that of the particle imbalance $I(t)$.
The results are nearly indistinguishable for system sizes $L>4$ for the 3D Anderson model and for system sizes $L>100$ for the 1D Aubry--André model.
In Figures \ref{figA1} and~\ref{figA2} of Appendix \ref{sec:differences}, we quantify the differences between ${I}(t)$ and $P^{}(t)$ and we show that the differences decrease with increasing the system size.

As a consequence of similarity between ${I}(t)$ and $P^{}(t)$ shown in Figure \ref{fig1}, both the rescaled survival probability~\cite{hopjan2023, hopjan2023scaleinvariant} and the rescaled particle imbalance~\cite{jiricek2023critical} can be used as indicators of critical behavior at the transition point.
This observation was the main result of \mbox{Refs.~\cite{hopjan2023, hopjan2023scaleinvariant, jiricek2023critical}}; in Appendix \ref{sec:scaleinvariance}, we summarize how the rescaling of the results in Figure~\ref{fig1} is carried out to detect the critical behavior.
We also show in Figure \ref{figA3} of Appendix \ref{sec:scaleinvariance} that the exponent of the power-law decay of $I(t)$ in the 1D Aubry--André model for typical initial states considered here matches the exponent of the power-law decay of $P(t)$.
This is not the case when considering atypical initial states such as the CDW states~\cite{jiricek2023critical}.

\begin{figure}[H]\vspace{-7pt}
\includegraphics[width=6.3 cm]{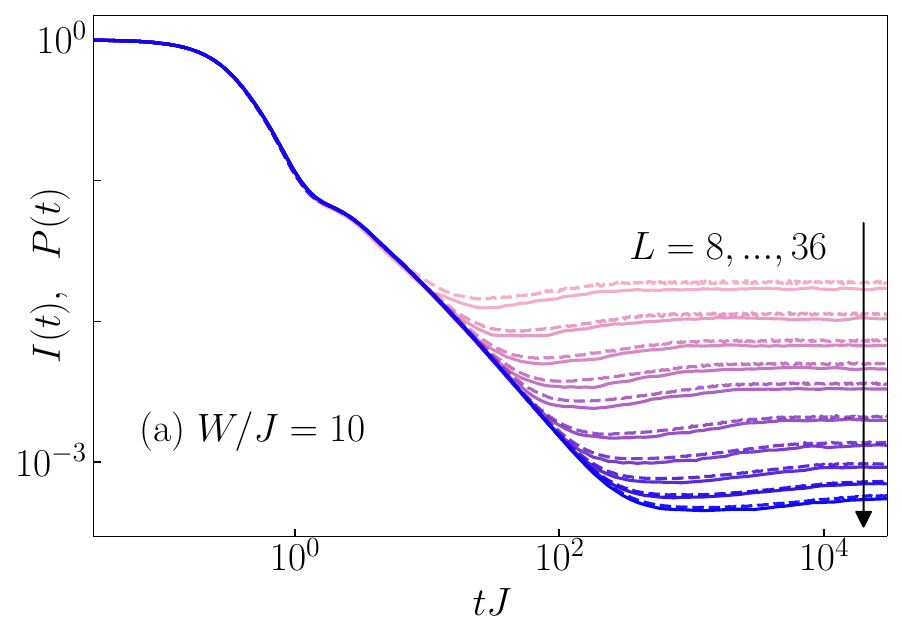}
\includegraphics[width=6.3 cm]{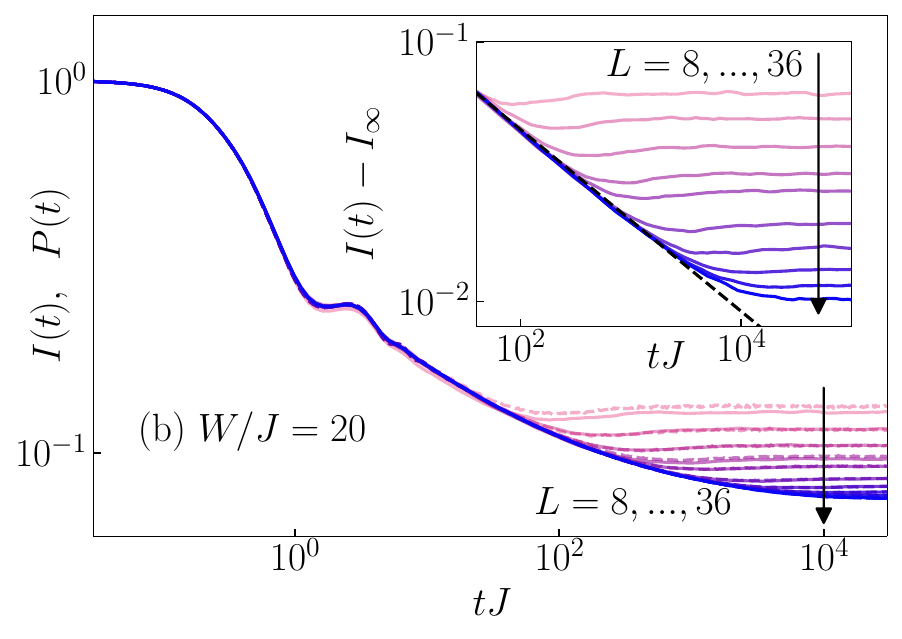}
\vspace{-0.2cm}
\caption{{
Dynamics of the survival probability $P(t)$ (dashed lines) and particle imbalance $I(t)$ (solid lines), see Equation (\ref{eq:equivalence5}), from the typical initial product states.
Results are shown for the 3D Anderson 
}}
\label{fig2a}
\end{figure}

\begin{figure}[H]
\includegraphics[width=6.5 cm]{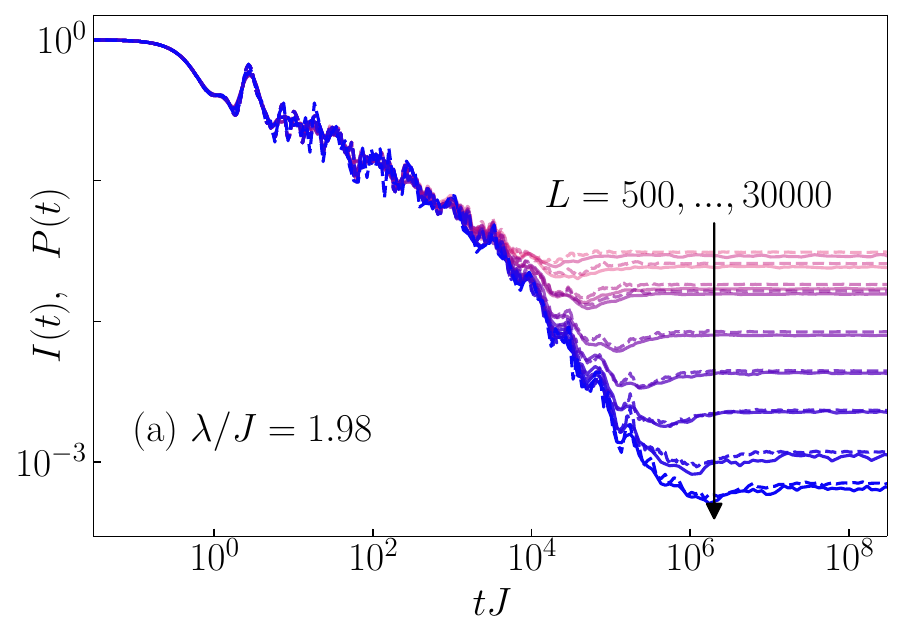}
\includegraphics[width=6.5 cm]{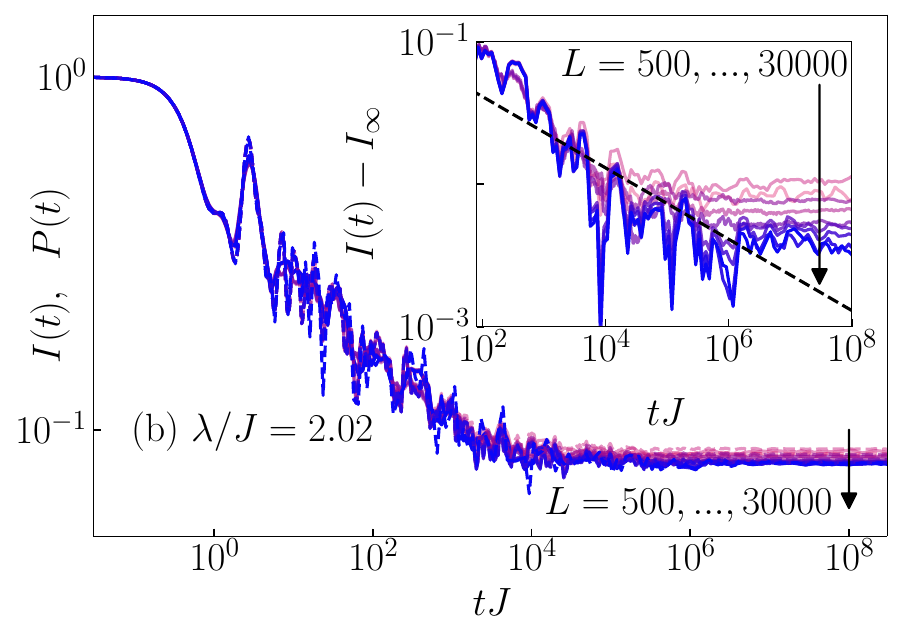}
\vspace{-0.2cm}
\caption{
{Dynamics} 
 of the survival probability $P(t)$ (dashed lines) and particle imbalance $I(t)$ (solid lines), see Equation (\ref{eq:equivalence5}), from the typical initial product states.
Results are shown for the 1D Aubry--André model (\textbf{a}) in the delocalized regime $\lambda/J = 1.98$ and (\textbf{b}) in the localized regime $\lambda/J=2.02$ for system sizes $L =$ 500, 750, 1000, 1250, 2500, 5000, 10,000, 20,000, 30,000. Inset of (\textbf{b}): subtracted imbalance $I(t)-I_\infty$, which reveals its power-law decay (dashed line). Here, $I_\infty$ is the infinite-time value of $I(t)$ in the thermodynamic limit.
}
\label{fig2}
\end{figure}

While the emergence of scale-invariant dynamics of rescaled quantities is limited to the critical point~\cite{hopjan2023, hopjan2023scaleinvariant, jiricek2023critical}, the similarity between the particle imbalance ${I}(t)$ and the survival probability $P^{}(t)$, our main result here, also emerges away from the critical point. {We} show evidence that the similarity between ${I}(t)$ and $P^{}(t)$ is not restricted to the eigenstate transition point of {the 3D Anderson or the} 1D Aubry--André model.
We compare ${I}(t)$ to $P^{}(t)$ in the delocalized regime; {specifically,
for $W/J=10$ in the 3D Anderson model, see Figure \ref{fig2a}a, and}
for $\lambda/J=1.98$ {in the 1D Aubry--André model}, see Figure \ref{fig2}a.
{We compare ${I}(t)$ to $P^{}(t)$ in the localized regime; specifically, $W/J=20$ in the 3D Anderson model, see Figure \ref{fig2a}b, and}
for $\lambda/J=2.02$ {in the 1D Aubry--André model}, see Figure \ref{fig2}b.
Even though {in the 1D Aubry--André model} the considered values of $\lambda/J$ are close to the critical point $\lambda_c/J=2$, they do not exhibit features of scale-invariant critical dynamics, as demonstrated in Figure \ref{figA4} of Appendix \ref{sec:scaleinvariance}.
We observe that the time evolution profile of $P^{}(t)$ is still very close to that of the imbalance $I(t)$ in both {the delocalized regimes, $W<W_c$ and} $\lambda<\lambda_c$, and {the localized regimes, $W>W_c$ and} $\lambda>\lambda_c$.
The agreement occurs despite the two regimes exhibiting distinct dynamical properties.
On the one hand, in the delocalized regime at {$W/J=10$ and} $\lambda/J=1.98$, the slope of the decay of $I(t)$ appears to get steeper with the increasing system size and the imbalance decays to zero in the thermodynamic limit, see {Figures \ref{fig2a}a and} \ref{fig2}a{, respectively}.
On the other hand, we observe in the localized regime the decay of $I(t)$ towards the infinite time value $\bar{I}$ that appears to saturate to a nonzero $I_\infty$ in the thermodynamic limit, see {Figures \ref{fig2a}b and} \ref{fig2}b.
The insets of \mbox{{Figures \ref{fig2a}b and} \ref{fig2}b} reveal that there is a power-law decay of $I(t)-I_\infty$ (we extract $I_\infty$ in the insets of \mbox{{ Figures \ref{figA4a}b and } \ref{figA4}b),} which is analogous to the decay of the survival probability $P(t)-P_\infty$$~$\cite{hopjan2023}. 

{\captionof*{figure}{(\textbf{a}) in the delocalized regime $W/J = 10$ and (\textbf{b}) in the localized regime $W/J = 20$ for system sizes $L = 8, 10, 12, 14, 16, 20, 24, 28, 32, 36$. Inset of (\textbf{b}): subtracted imbalance $I(t)-I_\infty$, which reveals its power-law decay (dashed line). Here, $I_\infty$ is the infinite-time value of $I(t)$ in the thermodynamic limit.}}
\vspace{6pt}
\unskip

%


\section{Transition Probabilities and Density Correlation Functions (Generalized Imbalance)}
\label{sec:imbalance_correlations}

In this section, we generalize the results of Section \ref{sec:imbalance} for the single-particle survival probability to the single-particle transition probabilities between lattice sites $i$ and $j$ at distance $d$$~$\cite{jiricek2023critical},\vspace{-6pt}
\begin{align}
    \label{eq:def_Pd_corr}
    P^{\,H,\,(d)}(t) = \Big\langle \,\sum_{i,|i-j|=d}P^{\,H}_{ij}(t)\,\Big\rangle_j\;,
\end{align}
where $\braket{\dots}_j$ denotes the average over all possible initial states $|j\rangle$, and the distance is defined as the minimal number of hops between the two sites, $|i-j|  \equiv ||\mathbf{r}_i - \mathbf{r}_{j}||_1$. Note that in the limit $d=0$, we recover the averaged survival probability from Equation \eqref{eq:def_Pd} since $P^{\,H,\,(0)}(t) \equiv P^{H}(t)$.

It was shown in Ref.$~$\cite{jiricek2023critical} that the rescaled transition probabilities exhibit scale-invariant critical dynamics that share certain similarities with the rescaled survival probability.
Hence, they can also be applied to detect the eigenstate transitions via quantum dynamics.
Here, we search for observables that exhibit similar time evolution profiles as the transition probabilities at $d>0$.

We argue that the observables of interest are the non-equal time and space density correlation functions at distance $d$,
\begin{equation} \label{eq:def_correl}
C^{\,H,\,(d)}_{}(t)= \frac{4}{V} \sum_j \sum_{i, |i-j|=d } \Big[\hat n^{\,H}_i(t)-1/2\Big] \Big[\hat n_j(0)-1/2\Big] \;,
\end{equation}
which can be thought of as the generalization of particle imbalance from Equation (\ref{eq:autocorrel}) since $C^{\,H,\,(0)}_{}(t) = I^H(t)$.
Hence, we refer to the observable $C^{\,H,\,(d)}_{}(t)$ in Equation (\ref{eq:def_correl}) as the generalized imbalance.

We expect that the time evolution of generalized imbalance, for the typical initial product states discussed in Section \ref{sec:imbalance}, is nearly indistinguishable from the time evolution of single-particle transition probabilities,
\begin{equation} \label{def_Cdt_Pdt}
    C^{\,H,\,(d)}_{}(t) \approx P^{\,H,\,(d)}_{}(t) \;.\vspace{-6pt}
\end{equation}
Equation~(\ref{def_Cdt_Pdt}) can be seen as the generalization of Equation (\ref{eq:equivalence4}).
The origin of the similarity between the generalized imbalance and transition probabilities is based on the same argument as the one invoked for the imbalance and survival probabilities in Section \ref{sec:imbalance}, i.e., on the self-averaging property of the local environment of lattice sites in the typical initial product states.

Below, we provide numerical evidence for the validity of Equation (\ref{def_Cdt_Pdt}). 
Specifically, we numerically compare the averages over the Hamiltonian realizations, giving rise to the relationship
\begin{align}
    \label{eq:def_Cd_av}
    C^{(d)}_{}(t) \approx P^{(d)}_{}(t) \;,\;\;\; {\rm with}\;\;\;
    C^{\,(d)}(t) = \braket{\,C^{\,H,\,(d)}(t) }_H \;, \;\;\;
    P^{\,(d)}(t) = \braket{\,P^{\,H,\,(d)}(t) }_H \;.\vspace{-6pt}
\end{align}
The averages $\langle ... \rangle_H$ over Hamiltonian realizations are defined analogously to those in Section \ref{sec:imbalance}.

\begin{figure}[H]
\hspace{-5pt}
\includegraphics[width=12 cm]{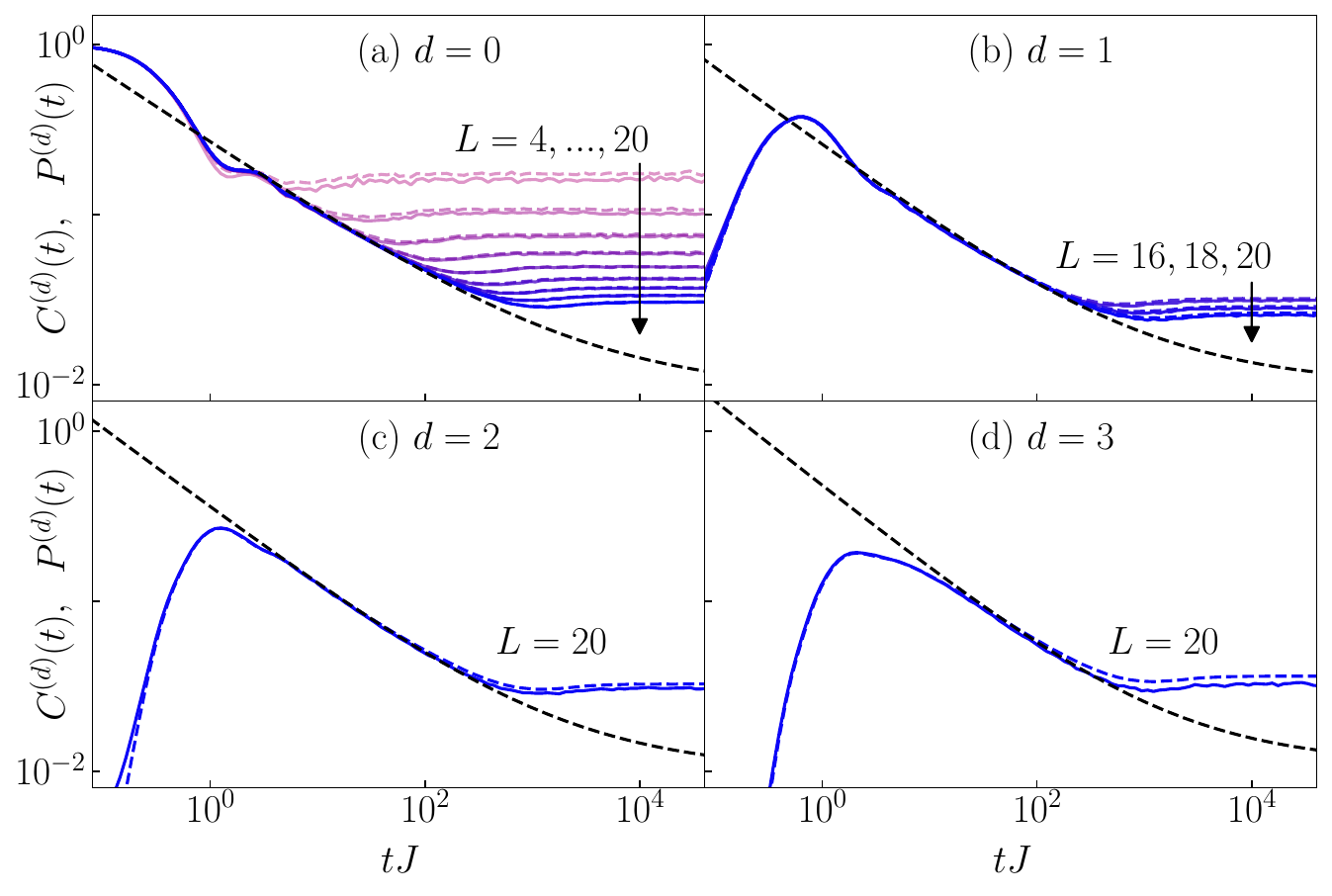}
\caption{Dynamics of the survival and transition probabilities $P^{(d)}(t)$ (dashed lines) and the corresponding generalized ~imbalance $C^{(d)}(t)$ ~(solid lines), see Equation (\ref{eq:def_Cd_av}), from the typical initial product states.
Results are shown for the 3D Anderson model at the critical point $W_c/J = 16.5$ for (\textbf{a}) $d = 0$ and $L = 4,6,8,10,12,14,16,18,20$, (\textbf{b}) $d = 1$ and $L = 16,18,20$, (\textbf{c}) $d = 2$ and $L = 20$, and \mbox{(\textbf{d}) $d = 3$} and $L = 20$. The black dashed line is a fit to the function $a_d (tJ)^{-\beta_d} + P^{(d)}_\infty$, where $P^{(d)}_\infty$ is the infinite-time value in the thermodynamic limit. 
}
\label{fig3}
\end{figure}

\begin{figure}[H]
\hspace{-5pt}
\includegraphics[width=12cm]{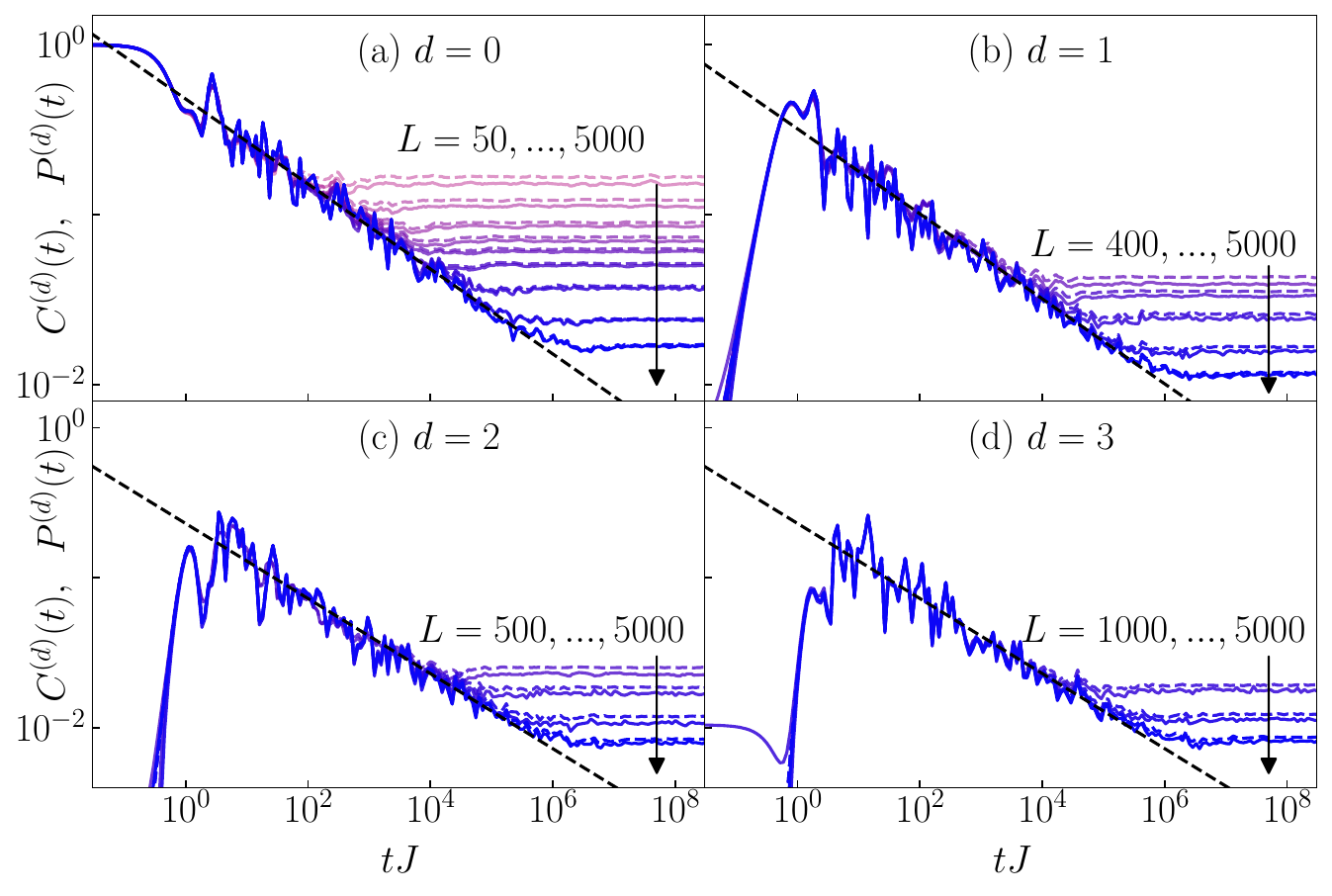}
\vspace{-0.2cm}
\caption{
Dynamics of the survival and transition probabilities $P^{(d)}(t)$ (dashed lines) and the corresponding generalized imbalance $C^{(d)}(t)$ (solid lines), see Equation (\ref{eq:def_Cd_av}), from the typical initial product states.
Results are shown for the 1D Aubry--André model at the critical point $\lambda_c/J = 2$  for \mbox{(\textbf{a}) $d = 0$} and $L = 50, 100, 200, 300, 400, 500, 1000, 2500, 5000$, (\textbf{b}) $d = 1$ and $L = 400, 500, 1000, 2500, 5000$, 
\mbox{(\textbf{c}) $d = 2$} and $L = 500, 1000, 2500, 5000$, and (\textbf{d}) $d = 3$ and $L = 1000, 2500, 5000$. The black dashed line is a fit to the function $a_d (tJ)^{-\beta_d}$.
}
\label{fig4}
\end{figure}

In Figures \ref{fig3} and$~$\ref{fig4}, we compare $C^{\,(d)}(t)$ with $P^{\,(d)}(t)$ for the 3D Anderson model and the 1D Aubry--André model, respectively, at their eigenstate transition points.
At $d=0$, we obtain the results from Section \ref{sec:imbalance} for the imbalance, $C^{\,(0)}(t)=I^{\,}(t)$, and the survival probability, $P^{\,(0)}(t)=P^{\,}(t)$---compare Figures \ref{fig3}a and \ref{fig4}a to  Figure \ref{fig1}a,b, respectively. 
At $d>0$, we indeed observe that the time evolution profile of the transition probability $P^{(d)}(t)$ is very close to that of the generalized imbalance $C^{\,(d)}(t)$. 
However, with increasing $d$, larger system sizes $L$ are required to observe the similarity.
Therefore, in \mbox{Figures \ref{fig3}b--d and \ref{fig4}b--d}, we only show results for the system sizes at which the agreement is reasonably high.
In Figures \ref{figA1} and~\ref{figA2} of Appendix \ref{sec:differences}, we quantify the differences between $C^{\,(d)}(t)$ and $P^{(d)}(t)$ and show that they decrease with the increasing system size.
In conclusion, the results from this section show that not only the particle imbalance but also the generalized imbalance that corresponds to non-equal time and space density correlations can be made, for typical initial product states, nearly indistinguishable from single-particle quantities, namely, the single-particle transition probabilities between different lattice sites.
\unskip
%


\section{Equal Time Connected Density--Density Correlation Functions}
\label{sec:correlations_equal}

So far, we have studied the dynamics of particle imbalance, which is a non-equal time density correlation function~(\ref{eq:autocorrel}), and the generalized imbalance, which is a non-equal time and space density correlation function~(\ref{eq:def_correl}).
We complement these studies by investigating another experimentally relevant quantity$~$\cite{Richerme14}, i.e., the equal-time connected density--density correlation function$~$\cite{Colmenarez19, Lezama_21, Lezama23, colbois2024interactiondriven}.
Even though the dynamics of the latter do not quantitatively agree with the dynamics of single-particle survival or transition probabilities, they still exhibit certain qualitative similarities that we discuss below.

The equal-time connected density--density correlation function is defined as the average of equal-time connected density--density correlations between the sites $i$ and $j$,
\begin{equation} \label{eq:correl_eq}
C^{\,H,\,(d)}_{eq}(t)= \frac{4}{V} \sum_j \sum_{i, |i-j|=d } C^{\,H,\,eq}_{ij}(t)\;,
\end{equation}
where the equal-time connected density--density correlation at sites $i$ and $j$ reads
\begin{align} \label{eq:correl_eq_ij}
C^{\,H,\,eq}_{ij}(t)=\langle \Psi_{t} | \Big[\hat{n}^{}_i-1/2\Big] \Big[\hat{n}^{}_j-1/2\Big] | \Psi_{t} \rangle-
\langle \Psi_{t} | \Big[\hat{n}^{}_i-1/2\Big] | \Psi_{t} \rangle\langle \Psi_{t} | \Big[\hat{n}^{}_j-1/2\Big] | \Psi_{t}\rangle
\;,
\end{align}
and $|\Psi_t\rangle = e^{-i\hat H t}|\Psi_0\rangle$.
Using Wick's theorem, one can split the density--density correlation term in Equation (\ref{eq:correl_eq_ij}) into two parts$~$\cite{Colmenarez19,Lezama23}, from which one of them cancels with the second term on the r.h.s.~of Equation (\ref{eq:correl_eq_ij}). The remaining term is the product of creation and annihilation operators at different sites, and it can be related to the elements of the one-particle density matrix
\begin{align} \label{eq:correl_eq_ij_1}
C^{\,H,\,eq}_{ij}(t)= \langle \Psi_{t} | \hat c^{\dagger}_i \hat c^{}_j | \Psi_{t} \rangle \langle \Psi_{t} | \hat c^{}_i \hat c^{\dagger}_j | \Psi_{t} \rangle= \rho^{\,H}_{ij}(t)\Big[\delta_{ij}-\rho^{\,H}_{ji}(t)\Big]\;,
\end{align}
where the time-dependent one-particle density matrix is defined as $\rho^{\,H}_{ij}(t)=\langle \Psi_{t} | \hat c^{\dagger}_i \hat c^{}_j | \Psi_{t} \rangle$. As a side remark, we note that the matrix elements of the one-particle density matrix can be expressed as
\begin{equation} \label{eq:def_rhoij_1}
\rho^{\,H}_{ij}(t) = \sum_{\substack{j_l \in \Psi_0}}^{} [G_{jj_l}^{\,H}(t)]^{*} G_{ij_l}^{\,H}(t)\;,
\end{equation}
where $G_{jj_l}^{\,H}=\langle j_l| e^{-i\hat Ht}|j\rangle$ is the propagator between states $|j\rangle$ and $|j_l\rangle$. 
One can interpret Equation \eqref{eq:def_rhoij_1} as the generalization of Equation \eqref{eq:def_ni_1}, since the latter reduces to the former at $i=j$. 

Before we proceed with the discussion of our numerical results, we first analyze Equation \eqref{eq:correl_eq_ij_1}. 
At $d=0$, the correlations $C^{\,H,\,eq}_{ii}(t)$ are non-negative since they can be expressed via the densities as $C^{\,H,\,eq}_{ii}(t)=n_i(t)-n_i^{2}(t)$.
Since $n_i(t) \in [0,1]$, it follows that $n_i(t)\geq n_i^{2}(t)$ and, hence, $C^{\,H,\,eq}_{ii}(t)\geq0$.
Moreover, the upper bound $C^{\,H,\,eq}_{ii}(t)\leq 1/4$ can be deduced from the same expression.
The two bounds then limit the values of the correlation function in Equation $~$\eqref{eq:correl_eq} at $d=0$ to the interval $[0,1]$.
On the other hand, at $d>0$, the correlations $C^{\,H,\,eq}_{ij}(t)$ are equal to $C^{\,H,\,eq}_{ij}(t)=-|\rho^{\,H}_{ij}(t)|^{2}$; hence, they are non-positive$~$\cite{Colmenarez19}, as is the correlation function in Equation \eqref{eq:correl_eq}. Finally, we also average $C^{\,H,\,(d)}_{eq}(t)$ over the Hamiltonian realizations,
\begin{align}
    \label{eq:def_Cd_eq_av}
    C^{\,(d)}_{eq}(t) = \braket{C^{\,H,\,(d)}_{eq}(t)}_H \;,
\end{align}
using the same protocol as in Sections \ref{sec:imbalance} and~\ref{sec:imbalance_correlations}.

\begin{figure}[H]
\includegraphics[width=6.5 cm]{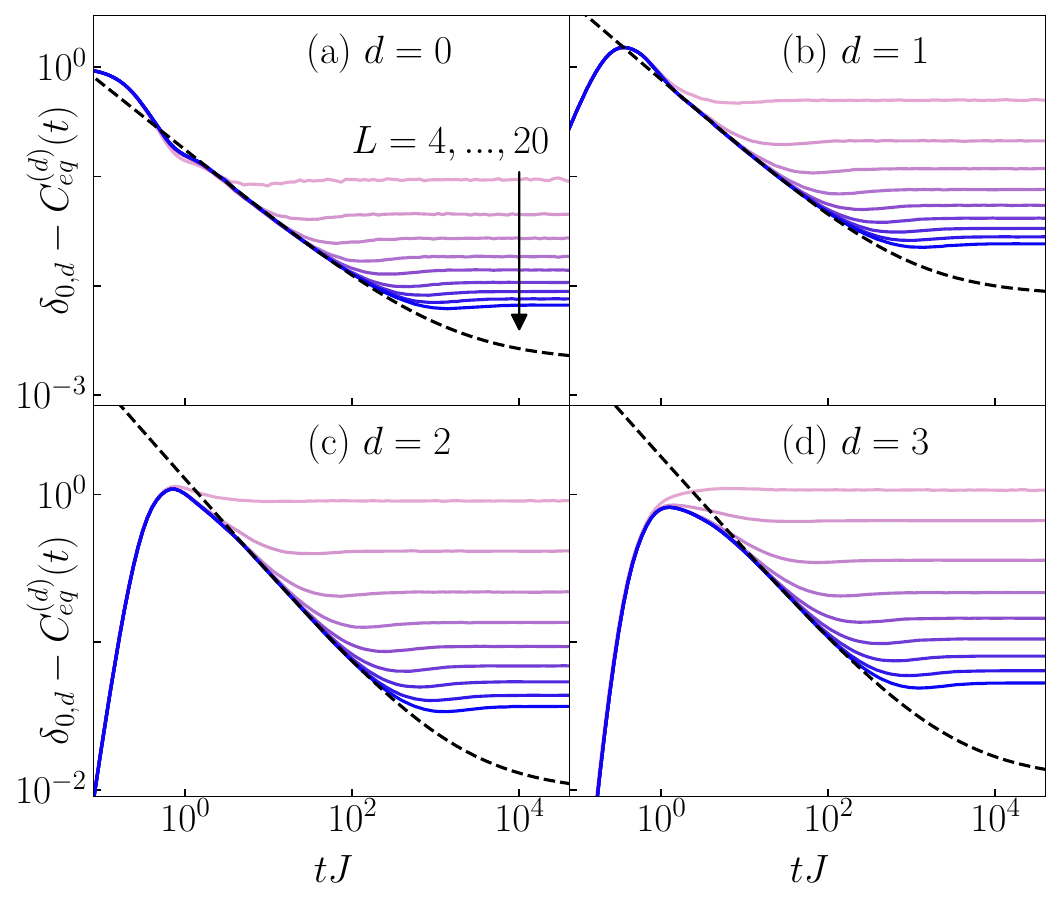}
\includegraphics[width=6.5 cm]{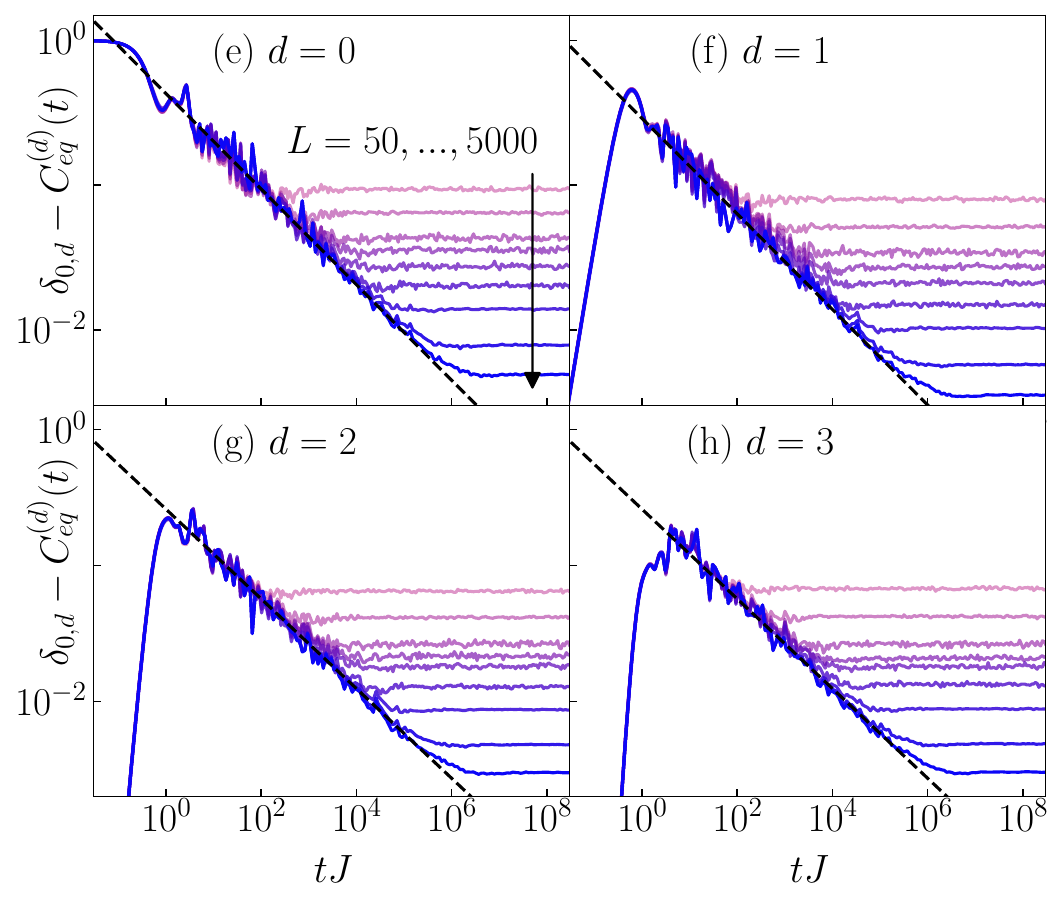}
\vspace{-0.2cm}
\caption{
Dynamics of equal-time connected density--density correlation functions  $\delta_{0,d}-C^{\,(d)}_{eq}(t)$, see Equation (\ref{eq:def_Cd_eq_av}), from the typical initial product states.
Results are shown for (\textbf{a}--\textbf{d}) the 3D Anderson model at the critical point $W_c/J = 16.5$ and for system sizes $L = 4,6,8,10,12,14,16,18,20$, and (\textbf{e}--\textbf{h}) the 1D Aubry--André model at the critical point $\lambda_c/J = 2$ and for system \linebreak sizes $L = 50, 100, 200, 300, 400, 500, 1000, 2500, 5000$. 
The black dashed line is a fit to the function $a_d (tJ)^{-\beta_d} + [\delta_{0,d}-C^{\,(d)}_{eq,\infty}]$, where $[\delta_{0,d}-C^{\,(d)}_{eq,\infty}]$ is the infinite-time value in the thermodynamic limit, see Appendix \ref{sec:scaleinvariance_2}.
}
\label{fig5}
\end{figure}

Based on the discussion above, in Figure$~$\ref{fig5}, we plot $1-C^{\,(d)}_{eq}(t)$ at $d=0$ and $-C^{\,(d)}_{eq}(t)$ at $d>0$, i.e., we plot $\delta_{0,d}-C^{\,(d)}_{eq}(t)$ instead of $C^{\,(d)}_{eq}(t)$ that can become negative.
Intriguingly, we observe qualitatively similar time evolution profiles as for the survival and transition probabilities.
The correlation function $1-C^{\,(0)}_{eq}(t)$ exhibits a power-law decay that is qualitatively similar to the decay of the survival probability, compare Figure \ref{fig5}a to \mbox{Figure \ref{fig3}a} and Figure \ref{fig5}e to Figure \ref{fig4}a. In the case of the 3D Anderson  model, $1-C^{\,(0)}_{eq}(t)$ decays towards a positive constant in the infinite system size limit, see Figure \ref{fig5}a, similarly to the decay of the survival probability to a non-zero constant $P_\infty$.
The correlation functions $-C^{\,(d)}_{eq}(t)$ at $d>0$ exhibit a maximum after which a power-law decay sets in, which is qualitatively similar to the behavior of the transition probabilities, compare Figure \ref{fig5}b--d to Figure \ref{fig3}b--d and Figure \ref{fig5}f--h to Figure \ref{fig4}b--d.
A closer inspection, however, reveals that the slopes of the decay of the equal-time connected density--density correlation functions are larger than in the case of the survival and transition probabilities; hence, in contrast to the results in Sections \ref{sec:imbalance} and~\ref{sec:imbalance_correlations}, a quantitative similarity does not emerge.

The qualitative similarity of the dynamics of the equal-time density--density correlation functions with the dynamics of the survival and transition probabilities motivates us to rescale the former analogously to the rescaling of the latter~\cite{hopjan2023, hopjan2023scaleinvariant}.
In \mbox{Appendices \ref{sec:scaleinvariance_2} and  \ref{sec:scaleinvariance_3}}, we show that, indeed, the rescaled equal-time connected density--density correlation functions exhibit scale-invariant mid-time and late-time dynamics, which are similar to the behavior of the survival and transition probabilities$~$\cite{hopjan2023, hopjan2023scaleinvariant, jiricek2023critical}.
Thus, the emergence of scale invariance in the dynamics of observables in many-body states appears to be a more general principle that does not necessarily require quantitative similarity with the dynamics of the survival and transition probabilities.

\section{Discussion}
\label{sec:discussion}

This work focuses on the dynamics of certain one-body observables in many-body states of quadratic fermionic models. 
We consider quantum quenches in which the initial states are many-body product states on a site occupation basis.
In contrast to Ref.~\cite{jiricek2023critical} that considered initial CDW states, we focused on typical product states with no order in particle occupations.
The main conclusions of this work are the following:

\begin{enumerate}
\item[(i)] We relate the dynamics of particle imbalance to the dynamics of single-particle survival probability, and we show that the two become nearly indistinguishable.
\item[(ii)] We extend the result (i) by showing that the generalized imbalance, i.e., the non-equal time and space density correlation function, also becomes nearly indistinguishable from the single-particle transition probabilities.
Results (i) and (ii) give a recipe for experiments on how to measure the properties of survival and transition probabilities using one-body observables.
\item[(iii)] We discuss the other experimentally relevant observables, i.e., the equal-time connected density--density correlation functions, which can be related to the one-particle density matrix observables.
We showed that these observables have qualitative, but not quantitative, similarities with the survival and transition probabilities. Importantly, they also appear to exhibit the scale-invariant dynamics at localization transitions; thus, they constitute an alternative route for the experimental observation of critical dynamics. 
\end{enumerate}

{Based on our analytical arguments, we expect that the main conclusions listed above are not limited to the two models (the 3D Anderson model and the 1D Aubry--André model) studied numerically in this work.}

As the final remark, we note that the most promising quantities for experiments, which allow for the detection of scale-invariant dynamics, are the particle imbalance and the equal-time connected density--density correlation function at distance $d = 0$. The particle imbalance allows one to measure both the scale-invariant critical dynamics and the fractal dimension of the underlying single-particle states but requires the measurement of both $n_i(0)$ and $n_i(t)$. On the other hand, the equal-time connected density--density correlation function at $d=0$ is a particularly simple quantity since it requires only the measurement of $n_i(t)$.

{As an outlook, we give another perspective on the main 
outcome of this work in Equations \eqref{eq:equivalence_half} and~\eqref{eq:equivalence4}.
We express the propagator $G^{\,H}_{jj}=\langle j| e^{-i\hat Ht}|j\rangle$ in Equation \eqref{eq:first} as a single-particle Green's function $G^{\,H}_{jj}=\langle\Psi_0| \hat c_j^\dagger(t)\hat c_j  |\Psi_0\rangle\ =\langle\Psi_0| e^{i\hat H t} \hat c_j^\dagger e^{-i\hat H t}  \hat c_j |\Psi_0\rangle$ evaluated in the initial many-body state $|\Psi_0\rangle$. Thus, we reformulate Equation \eqref{eq:equivalence_half} as\vspace{-6pt}
\begin{align}  \label{eq:equivalence_outlook}
    I_{}^{\,H}(t) \approx \frac{2}{L} \sum_{j\in\Psi_0}
    |\langle\Psi_0| \hat c_j^\dagger(t)\hat c_j  |\Psi_0\rangle|^2\,,
\end{align}
which is an expression based purely on single-particle observables in the many-body state. This formulation may motivate future studies, which should explore the validity of Equation \eqref{eq:equivalence_outlook} for interacting systems that contain non-quadratic terms in the Hamiltonian.}

\vspace{6pt} 

\authorcontributions{The computer codes were written and data produced by M.H. Both authors contributed to the data analysis and writing of the paper. All authors have read and agreed to the published version of the manuscript.}

\funding{{This} 
 work was supported by the Slovenian Research and Innovation Agency (ARIS), Research core fundings Grants No. P1-0044, N1-0273 and J1-50005.
We gratefully acknowledge the High
Performance Computing Research Infrastructure Eastern
Region (HCP RIVR) consortium~\cite{vega1} and
European High Performance Computing Joint Undertaking (EuroHPC JU)~\cite{vega2}  for funding
this research by providing computing resources of the
HPC system Vega at the Institute of Information Sciences~\cite{vega3}.}

\institutionalreview{{ Not applicable.}}

\dataavailability{All data are available upon request.}

\acknowledgments{{We} 
 acknowledge discussions with F. Heidrich-Meisner, S. Jiricek, and P. \L yd\.{z}ba.}

\conflictsofinterest{The authors declare no conflicts of interest.}

\appendixtitles{yes} 
\appendixstart
\appendix


\section{System Size Dependence of Differences between $P^{(d)}$ and $C^{(d)}$}
\label{sec:differences}

Here, we quantify the differences between the transition probabilities $P^{(d)}$ and the generalized imbalances $C^{(d)}$ from Equation \eqref{eq:def_Cd_av}, defined by their absolute differences,
\begin{align}
    \label{eq:def_diff}
    \sigma^{(d)}_{\rm abs}(t)=|C^{\,(d)}(t)-P^{\,(d)}(t)| \;.
\end{align}
At $d=0$, Equation (\ref{eq:def_diff}) reduces to the differences between the survival probabilities $P=P^{(0)}$ and the particle imbalance $I=C^{(0)}$, as discussed in Section \ref{sec:imbalance}.

\begin{figure}[H]
\includegraphics[width=6.5 cm]{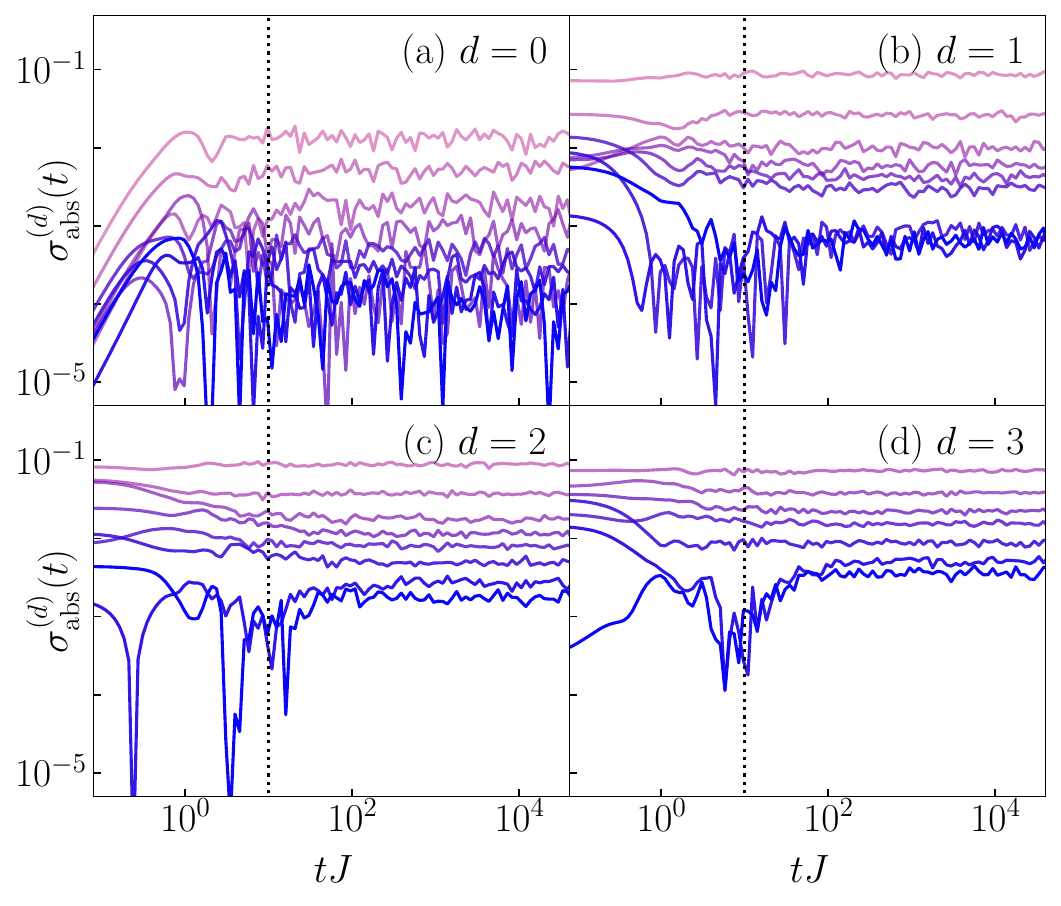}
\includegraphics[width=6.5 cm]{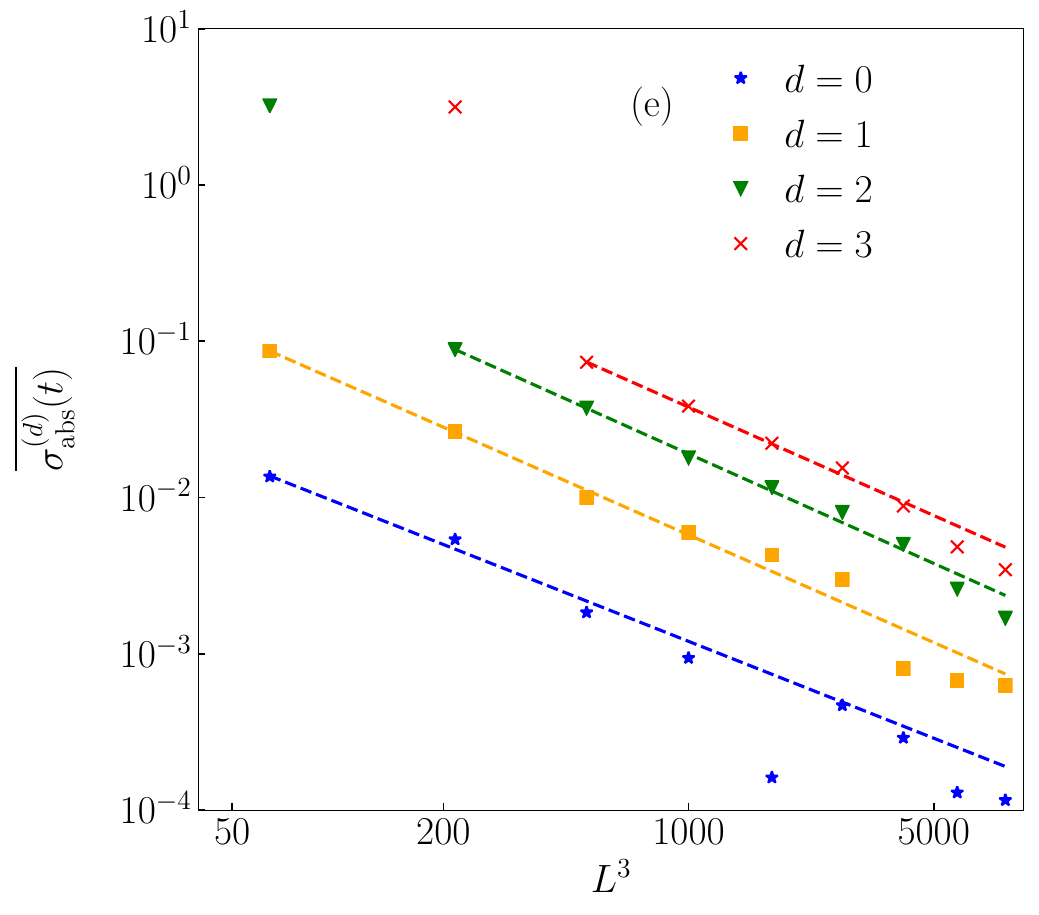}
\caption{The absolute differences $\sigma^{(d)}_{\rm abs}(t)$ for (\textbf{a}) $d = 0$, (\textbf{b}) $d = 1$, (\textbf{c}) $d = 2$, and (\textbf{d}) $d = 3$ for the 3D Anderson model at the critical point $W_c/J = 16.5$ and for system sizes $L = 4,6,8,10,12,14,16,18,20$, corresponding to the results in Figures \ref{fig1}a and \ref{fig3}. Panel (\textbf{e}) displays the averaged absolute differences $\overline{\sigma^{(d)}_{\rm abs}(t)}$ as a function of system volume $L^{3}$. The averaging is performed over times larger than $tJ = 10$ (dotted vertical lines in panels (\textbf{a}--\textbf{d})). The dashed lines in panel (\textbf{e}) are fits to the results of form $aL^{-b}$.}
\label{figA1}
\end{figure}

\begin{figure}[H]
\includegraphics[width=6.5 cm]{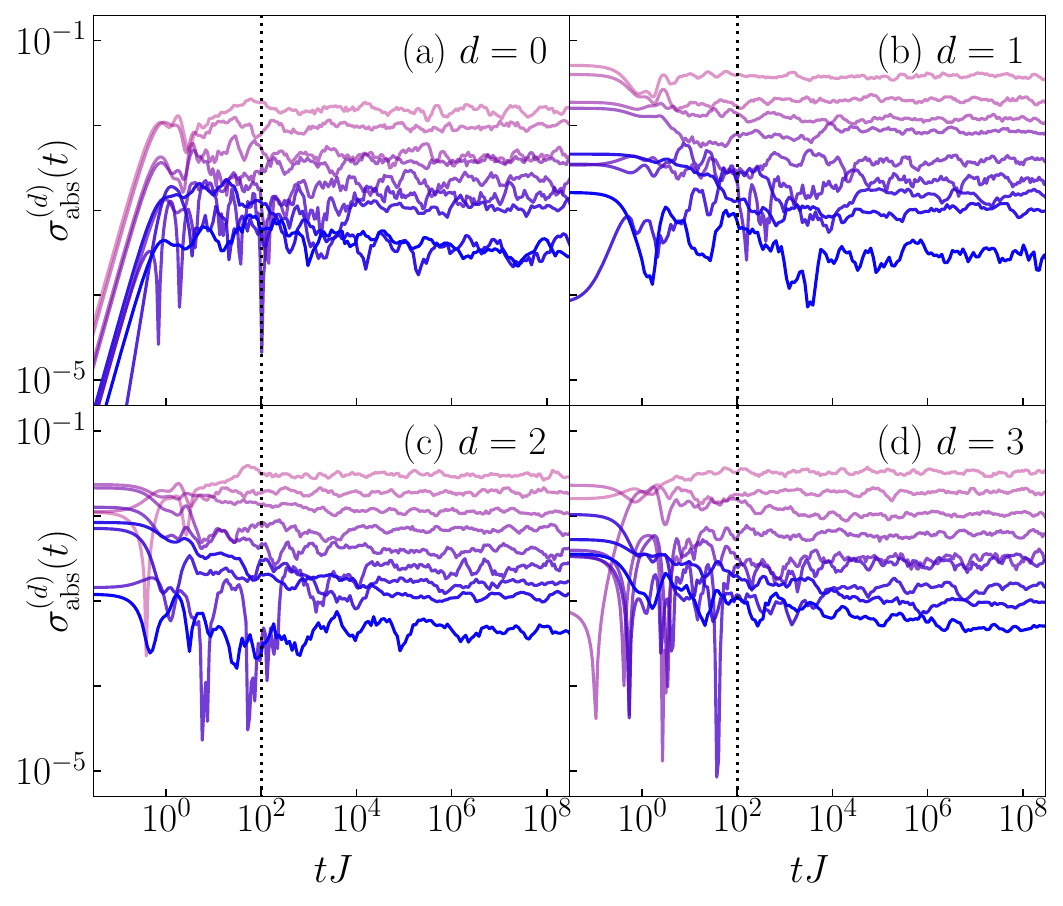}
\includegraphics[width=6.5 cm]{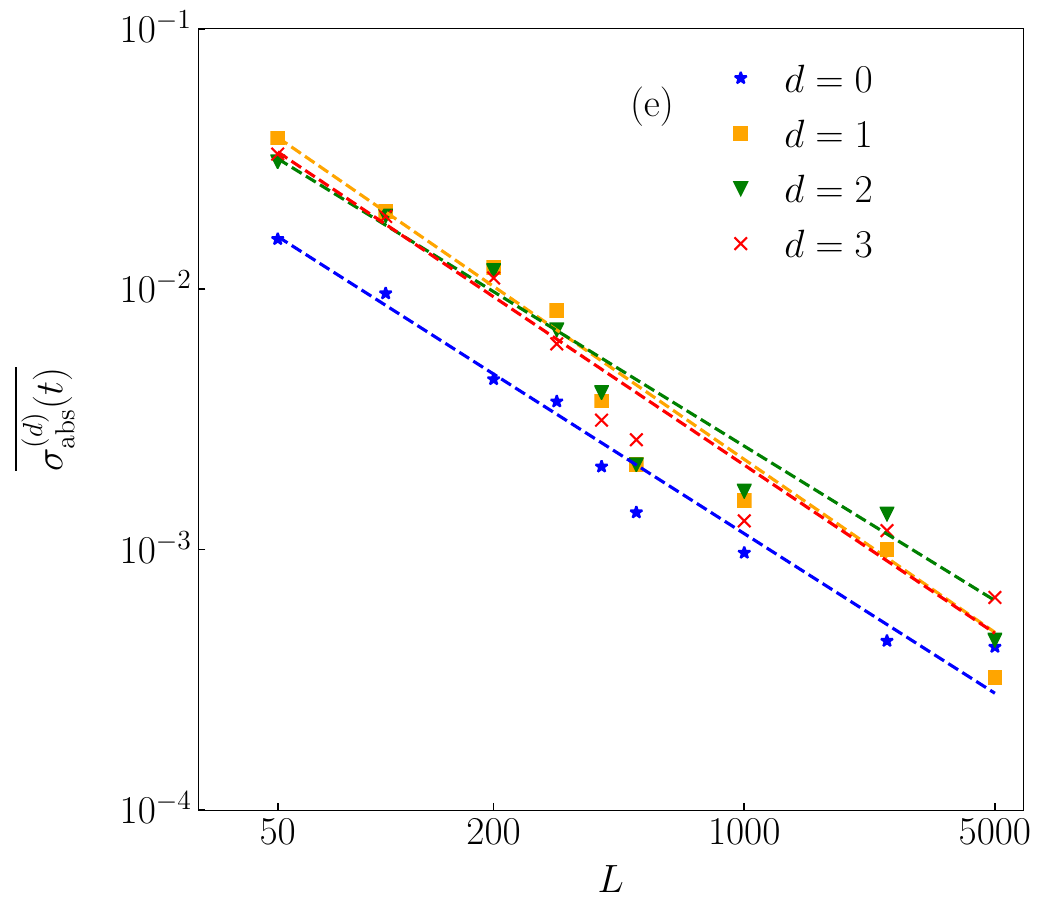}
\vspace{-0.2cm}
\caption{The absolute differences $\sigma^{(d)}_{\rm abs}(t)$ for (\textbf{a}) $d = 0$, (\textbf{b}) $d = 1$, (\textbf{c}) $d = 2$, and \mbox{(\textbf{d}) $d = 3$} for the 1D Aubry--André model at the critical point $\lambda_c/J = 2$ and for system sizes \linebreak $L = 50, 100, 200, 300, 400, 500, 1000, 2500, 5000$, corresponding to the results in Figures \ref{fig1}b and \ref{fig4}. Panel (\textbf{e}) displays the averaged absolute differences $\overline{\sigma^{(d)}_{\rm abs}(t)}$ as a function of system size $L$. The averaging is performed over times larger than $tJ = 10^{2}$ (dotted vertical lines in panels (\textbf{a}--\textbf{d})). The dashed lines in panel (\textbf{e}) are fits to the results of form $aL^{-b}$.
}
\label{figA2}
\end{figure}

In Figures \ref{figA1}a--d and \ref{figA2}a--d, we show $\sigma^{(d)}_{\rm abs}(t)$ for the 3D Anderson model and the 1D Aubry--André model, respectively. We observe a decrease in the differences with the system size. To quantify the system size dependence of the differences, we first perform their time averages defined as
\begin{align}
    \label{eq:def_diff_averaged}
    \overline{\sigma^{(d)}_{\rm abs}(t)}=\frac{1}{N_t}\sum_{\{t_i\}}\sigma^{(d)}_{\rm abs}(t_i)\;,
\end{align}
\textls[-15]{where $N_t$ is the number of values in the discrete time set $\{t_i\}$. For the averages, we consider only times larger than $tJ=10^{1}$ and $tJ=10^{2}$ for the 3D Anderson model and the 1D Aubry--André model, respectively, at which the differences approach their steady-state values. 
The vertical dotted lines in Figures \ref{figA1}a--d and \ref{figA2}a--d  denote the onset of the steady-state behavior.}

In Figures \ref{figA1}e and \ref{figA2}e, we plot $\overline{\sigma^{(d)}_{\rm abs}(t)}$ as a function of the number of lattice sites for the 3D Anderson model and the 1D Aubry--André model, respectively.
The results are consistent with a power-law decay. We perform a fit to the results of form $aL^{-b}$, where $a$ and $b$ are fitting parameters. For the 3D Anderson model, we find $b\approx3$, and for the 1D Aubry--André model, we find $b\approx1$, which suggests that in both cases the differences vanish upon increasing the system size.

 \vspace{-12pt}

%


\section{Rescaled Imbalance and Transition Probabilities: Scale-Invariant Dynamics at Eigenstate Transitions}
\label{sec:scaleinvariance}

\begin{figure}[H]
\hspace{-5pt}
\includegraphics[width=6.3 cm]{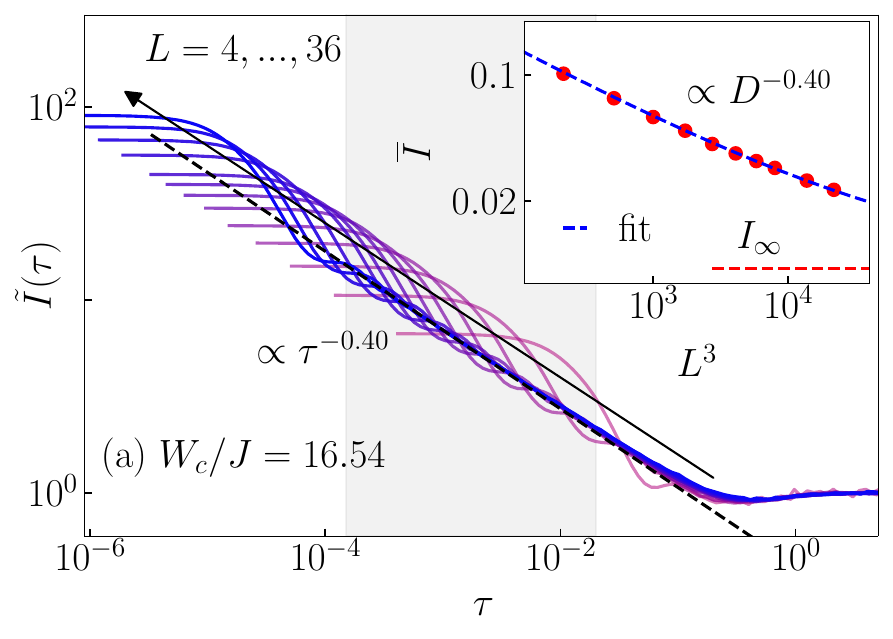}
\includegraphics[width=6.3 cm]{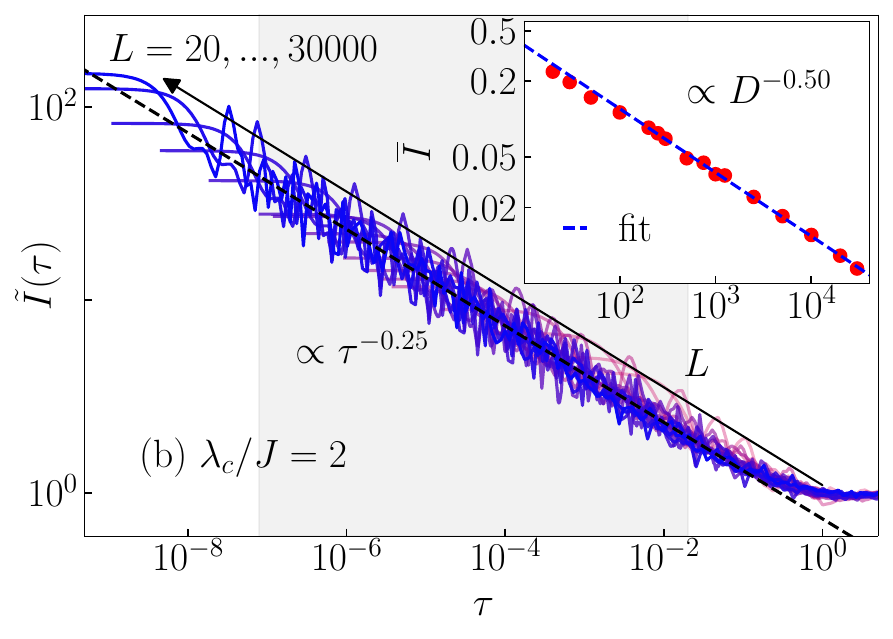}
\caption{
{Dynamics} 
 of the rescaled particle imbalance $\tilde{I}(\tau)$ as a function of scaled time $\tau$, see Equation (\ref{eq:imbalance_rescaled}).
Data are the same as in Figure~\ref{fig1}, and they are shown for (\textbf{a}) the 3D Anderson model at the critical point $W_c/J=16.5$ and for system sizes $L = 4,6,8,10,12,14,16,18,20,24,28,32,36$, and (\textbf{b}) the 1D Aubry--André model at the critical point $\lambda_c/J=2$ and for system sizes \linebreak $L$ = 20, 30, 50, 100, 200, 250, 300, 400, 500, 750, 1000, 1250, 2500, 5000, 10,000, 20,000, 30,000.
The dashed line is a two-parameter fit (within the shaded regions) to the function $\tilde{I}(\tau) = a_I\,\tau^{-\beta_I}$.
Inset: Infinite-time values $\Bar{I}$ versus the single-particle Hilbert-space dimension $D=L^3$ (circles) and the three-parameter fits to Equation (\ref{eq:I_scaling}) (dashed line). The horizontal dashed line is the infinite-time value in the thermodynamic limit $I_\infty$.
}
\label{figA3}
\end{figure}

\begin{figure}[H]
\hspace{-5pt}
\includegraphics[width=6.3 cm]{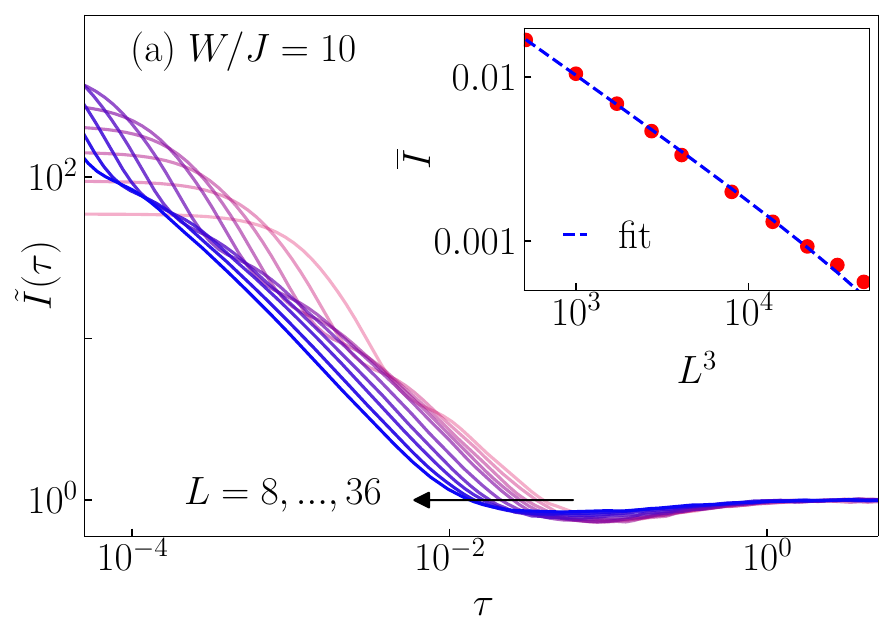}
\includegraphics[width=6.3 cm]{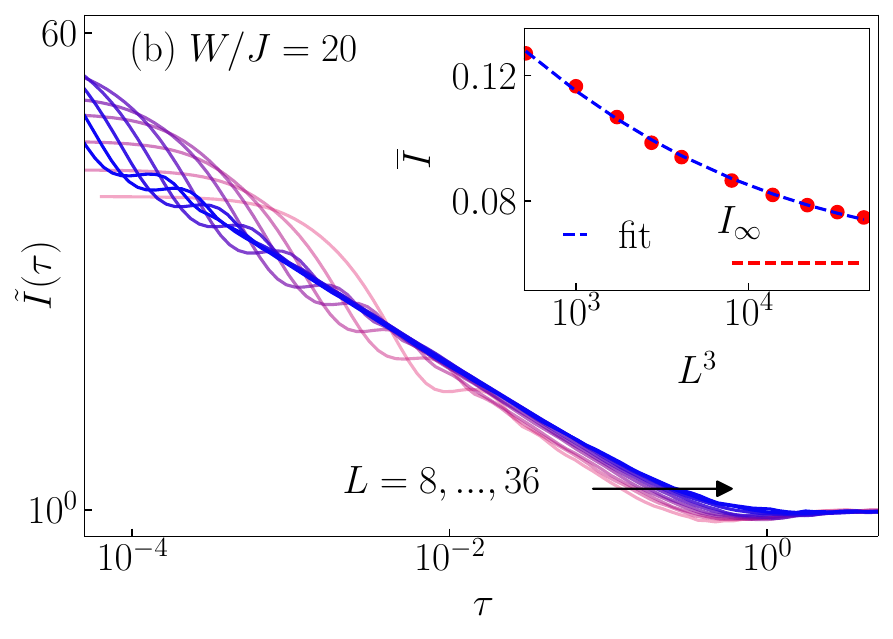}
\caption{{
Dynamics of the rescaled particle imbalance $\tilde{I}(\tau)$ as a function of scaled time $\tau$, see Equation (\ref{eq:imbalance_rescaled}).
Data are the same as in Figure~\ref{fig2a}, and they are shown for the 3D Anderson model (\textbf{a}) in the delocalized regime $W/J=10$ and {(\textbf{b})} 
 in the localized regime $W/J=20$ for system sizes $L = 8, 10, 12, 14, 16, 20, 24, 28, 32, 36$.
Inset: Infinite-time values $\Bar{I}$ versus the single-particle Hilbert-space dimension $D=L^{3}$ (circles) and the three-parameter fits to Equation (\ref{eq:I_scaling}) (dashed line). 
The horizontal dashed line is the infinite-time value in the thermodynamic limit $I_\infty$.
}}
\label{figA4a}
\end{figure}

\begin{figure}[H]
\hspace{-4pt}
\includegraphics[width=6.3 cm]{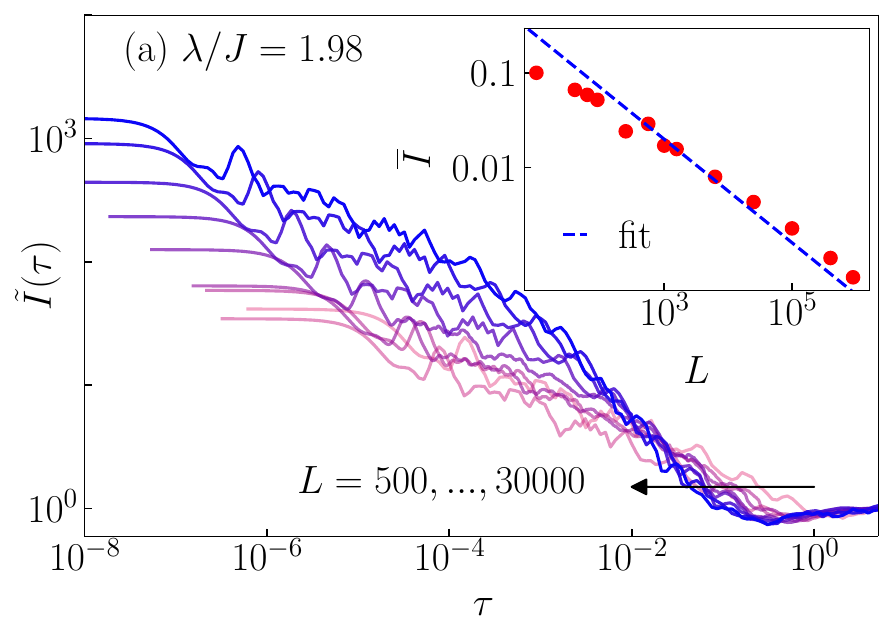}
\includegraphics[width=6.3 cm]{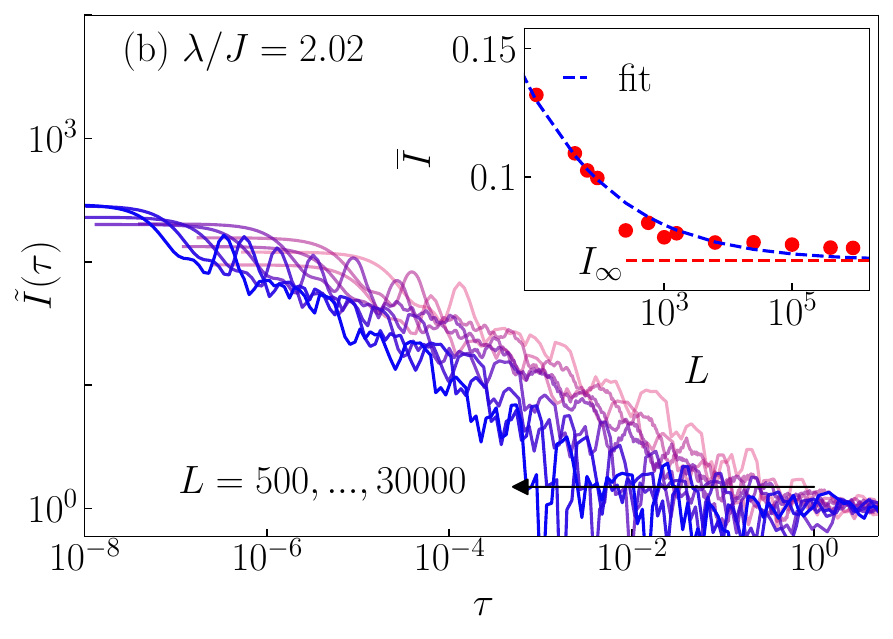}
\caption{
{Dynamics} 
 of the rescaled particle imbalance $\tilde{I}(\tau)$ as a function of scaled time $\tau$, see Equation (\ref{eq:imbalance_rescaled}).
Data are the same as in Figure~\ref{fig2}, and they are shown for the 1D Aubry--André model (\textbf{a}) in the delocalized regime $\lambda/J=1.98$ and (\textbf{b}) in the localized regime $\lambda/J=2.02$ for system sizes $L$ = 500, 750, 1000, 1250, 2500, 5000, 10,000, 20,000, 30,000.
Inset: Infinite-time values $\Bar{I}$ versus the single-particle Hilbert-space dimension $D=L$ (circles) and the three-parameter fits to Equation (\ref{eq:I_scaling}) (dashed line). 
The horizontal dashed line is the infinite-time value in the thermodynamic limit $I_\infty$.
}
\label{figA4}
\end{figure}

Motivated by Refs.$~$\cite{hopjan2023, hopjan2023scaleinvariant,jiricek2023critical}, 
we perform rescaling of the results for particle imbalance in Figures \ref{fig1}--\ref{fig2}.
As the dynamics of particle imbalance are nearly indistinguishable from the dynamics of survival probability, 
the originally proposed rescaling for the survival probability in Refs.$~$\cite{hopjan2023, hopjan2023scaleinvariant} reduces to the rescaling 
of the imbalance  proposed in Ref.$~$\cite{jiricek2023critical}. The rescaled imbalance$~$\cite{jiricek2023critical} is defined as 
\begin{align}\label{eq:imbalance_rescaled}
    \tilde{I}(\tau) = \frac{I(\tau) - I_\infty}{\overline{I} - I_\infty} \;,
\end{align}
where the infinite-time value at a fixed system size reads $\overline{I}^{} = \lim_{t\to\infty} I^{}(t)$, and the corresponding value in the thermodynamic limit $I_\infty$ is extracted from the power-law decay ansatz,
\begin{equation} \label{eq:I_scaling}
\overline{I}=c_I\,D^{-\gamma_I}+I_\infty \;,
\end{equation}
where $D = V$ is the single-particle Hilbert space dimension.
Fits of the results to this ansatz are shown in the insets of Figures \ref{figA3}--\ref{figA4}.
The scaled time $\tau$ is measured in units of the typical Heisenberg time $t_H^\mathrm{typ}$,
\begin{align} \label{def_tHeis}
    \tau=t/t_H^\mathrm{typ}\;,\;\;\; t_H^\mathrm{typ} = 2\pi\,\text{e}^{-\braket{\braket{\ln(\varepsilon_{q+1} - \varepsilon_q)}_q}_H} \;,
\end{align}
in which $\braket{\dots}_q$ denotes the average over all neighboring single-particle eigenenergies $\varepsilon_q$ of $\hat{H}$ and $\braket{\dots}_H$ is the average over Hamiltonian realizations.

In Figure \ref{figA3}a,b, we plot the rescaled imbalance $\tilde{I}(\tau)$ for the 3D Anderson model and the 1D Aubry--André model at their critical points, respectively. The data in Figure \ref{figA3} are identical to those in Figure \ref{fig1}. We observe clear emergence of scale invariance in the critical dynamics of the particle imbalance, which is consistent with the results for the survival probability in Ref.~\cite{hopjan2023}.

\unskip

We note that the scale-invariant dynamics at the critical point was also shown to emerge for the transition probabilities $P^{\,(d)}(t)$ from Equation \eqref{eq:def_Cd_av}, see Ref.~\cite{jiricek2023critical}. 
Hence, since the generalized imbalance $C^{\,(d)}(t)$ from Equation \eqref{eq:def_Cd_av} becomes nearly indistinguishable from the transition probability $P^{\,(d)}(t)$, we expect that $C^{\,(d)}(t)$ ultimately develops the scale invariance as well.
However, this is only expected to occur for very large system sizes; therefore, the generalized imbalances $C^{\,(d)}(t)$ for $d>0$ are less useful for the experimental detection of scale invariance than the particle imbalance. Thus, we do not explicitly discuss here the rescaling of $C^{\,(d)}(t)$ for $d>0$.

In Figures \ref{figA4a} and~\ref{figA4}, we show the rescaled imbalance $\tilde{I}(\tau)$ in the {3D Anderson model and} 1D Aubry--André model{, respectively,}  at the two disorder strengths away from the critical point.
{The data in Figures \ref{figA4a}a and \ref{figA4}a are identical to those in \mbox{Figures \ref{fig2a}a and \ref{fig2}a}, respectively, and the data in Figures \ref{figA4a}b and \ref{figA4}b are identical to those in \mbox{Figures \ref{fig2a}b and \ref{fig2}b}, respectively.
In all} cases, we do not observe indications of  scale invariance. 
{In the 3D Anderson model at $W/J=10$, the decay as a function of $\tau$ becomes faster with increasing $L$, see Figure~\ref{figA4a}a, which complies with the relaxation time shorter than the typical \mbox{Heisenberg time~\cite{suntajs_prosen_21}}. Conversely, at $W/J=20$, the relaxation time is larger than the Heisenberg time~\cite{suntajs_prosen_21}, which results in the opposite drifts of the curves with the system size~\cite{jiricek2023critical}, see Figure~\ref{figA4a}b.
In the 1D Aubry--André model at} $\lambda/J=1.98$, the decay as a function of $\tau$ becomes faster with increasing $L${, see Figure~\ref{figA4}a}, again indicating relaxation at times shorter than the typical Heisenberg time. At $\lambda/J=2.02$, the relaxation also occurs at times shorter than the typical Heisenberg time{, see Figure~\ref{figA4}b}, which causes the scaled imbalance $\tilde{I}(\tau)$ to shift to the left with increasing $L$.
Hence, while the rescaled imbalance exhibits scale-invariant critical dynamics at the eigenstate transition points, this does not appear to be the case away from it.

\section{Equal-Time Connected Density--Density Correlation Functions: Scale-Invariant Dynamics at Eigenstate Transitions}
\label{sec:scaleinvariance_2}

\begin{figure}[H]
\hspace{-5pt}
\includegraphics[width=12 cm]{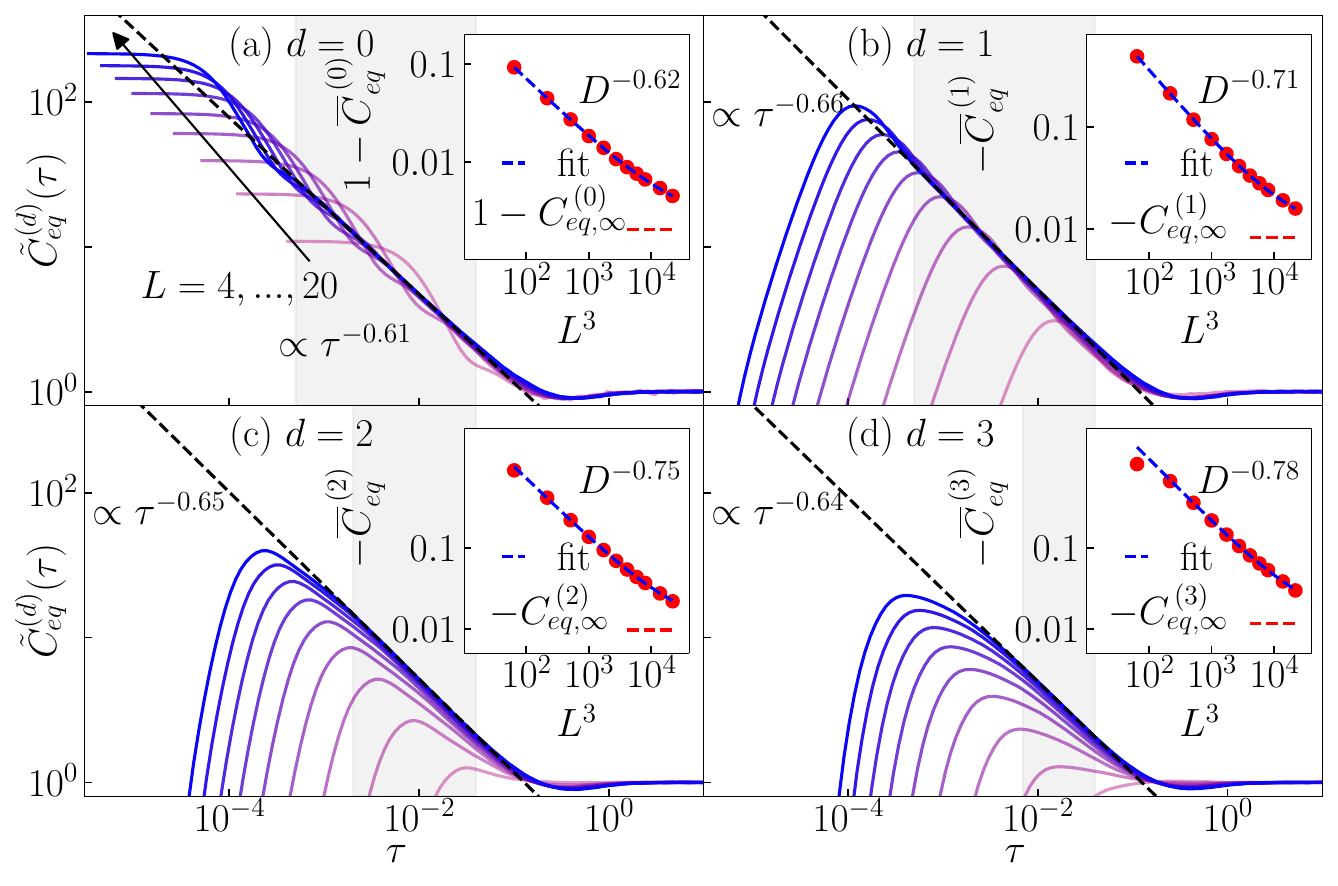}
\caption{
Dynamics of the rescaled equal-time connected density--density correlation functions  $\tilde{C}^{\,(d)}_{eq}(\tau)$, see Equation (\ref{eq:correl_equal_rescaled1}), from the typical initial product states.
Results are shown for the 3D Anderson model at the critical point $W_c/J=16.5$ and for system sizes $L = 4,6,8,10,12,14,16,18,20$. 
The dashed line is a two-parameter fit (within the shaded regions) to the power-law decay from\mbox{ Equation (\ref{def_Ceq_powerlaw})}.
Inset: Infinite-time values $\delta_{0,d}-\overline{C}^{\,(0)}_{eq}$ versus the single-particle Hilbert-space dimension $D=L^3$ (circles) and the three-parameter fits to Equation (\ref{eq:correlations_scaling}) (dashed line). The horizontal dashed line is the infinite-time value in the thermodynamic limit $\delta_{0,d}-C^{\,(d)}_{eq,\infty}$.
}
\label{figA5}
\end{figure}

\begin{figure}[H]
\hspace{-5pt}
\includegraphics[width=12 cm]{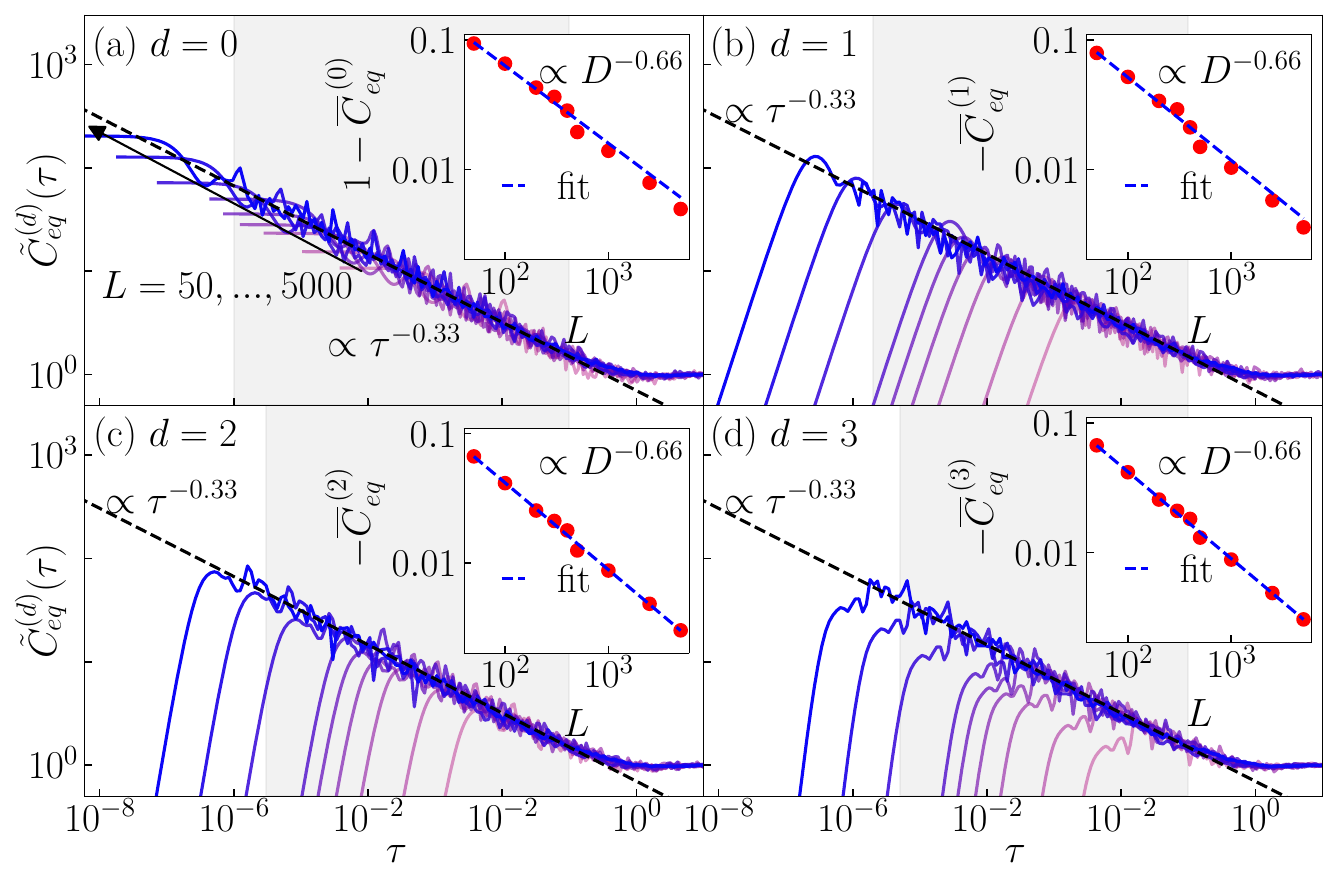}
\vspace{-0.2cm}
\caption{
Dynamics of the rescaled equal-time connected density--density correlation functions  $\tilde{C}^{\,(d)}_{eq}(\tau)$, see Equation (\ref{eq:correl_equal_rescaled1}), from the typical initial product states.
Results are shown for the 1D Aubry--André model at the critical point $\lambda_c/J=2$ and for system sizes \linebreak $L = 50, 100, 200, 300, 400, 500, 1000, 2500, 5000$. 
The dashed line is a two-parameter fit (within the shaded regions) to the power-law decay from Equation (\ref{def_Ceq_powerlaw}).
Inset: Infinite-time values $\delta_{0,d}-\overline{C}^{\,(0)}_{eq}$ versus the single-particle Hilbert-space dimension $D=L$ (circles) and the three-parameter fits to Equation (\ref{eq:correlations_scaling}) (dashed line). 
}
\label{figA6}
\end{figure}

Interestingly, even though the dynamics of the equal-time connected density--density correlation functions $C^{\,(d)}_{eq}(t)$ are not quantitatively related to the dynamics of the survival and transition probabilities, when rescaled, they develop scale invariance analogous to those discussed in Appendix \ref{sec:scaleinvariance}. 
Indeed, it was argued in Ref.~\cite{jiricek2023critical} that the scale invariance at the eigenstate transition is not limited to the observables that become indistinguishable from the survival or transition probabilities; hence, the emergence of scale invariance can be considered as a general principle. Here, we follow this principle and we perform a rescaling, analogous to the rescaling in Equation (\ref{eq:imbalance_rescaled}), of the equal-time connected density--density correlations functions from Section \ref{sec:correlations_equal}. The rescaling is defined as
\begin{align}\label{eq:correl_equal_rescaled1}
    \tilde{C}^{\,(d)}_{eq}(\tau) 
    = \frac{C^{\,(d)}_{eq}(\tau) -C^{\,(d)}_{eq,\infty}}{\overline{C}^{\,(d)}_{eq} - C^{\,(d)}_{eq,\infty}} \;,
\end{align}
where the infinite-time values read $\overline{C}^{\,(d)}_{eq} = \lim_{t\to\infty} C^{\,(d)}_{eq}(t)$. The thermodynamic limit of $\overline{C}^{\,(d)}_{eq} $ is extracted from the power-law decay ansatz
\begin{equation} \label{eq:correlations_scaling}
\overline{C}^{\,(d)}_{eq} =c_C^{\,(d)}\,D^{-\gamma_C^{\,(d)}}+C^{\,(d)}_{eq,\infty} \;. 
\end{equation}
In Figures \ref{figA5}a--d and \ref{figA6}a--d, we plot the rescaled equal-time connected density--density correlation functions $\tilde{C}^{\,(d)}_{eq}(\tau)$ for the 3D Anderson model and the 1D Aubry--André model at their critical points, respectively. We observe the emergence of scale invariance in all panels.

%

Another property that can be observed from Figure~\ref{figA5} is that the decay follows a power-law,
\begin{equation} \label{def_Ceq_powerlaw}
    \tilde{C}^{\,(d)}_{eq}(\tau) = a_C^{(d)} \tau^{-\beta_C^{(d)}}.
\end{equation}
The power-law exponent $\beta_C^{\,(d)}$ is generally larger than the corresponding exponent for the transition probability $\beta_d$~\cite{jiricek2023critical}. In the case of the 3D Anderson model, the values of $\beta_d$ for $d \in \{0,1,2,3\}$ are in range $\beta_d \in [0.4,0.45]$~\cite{jiricek2023critical} and the corresponding values of $\beta_C^{\,(d)}$ in Figure \ref{figA5}a--d are in range $\beta_C^{\,(d)}\in [0.61,0.66]$.
It remains an open question as to whether the variations in values of $\beta_C^{\,(d)}$ for different $d$ are caused by finite size effects, converging ultimately to the same value in the thermodynamic limit, or they will remain distinct. 
In the 1D Aubry--André model, the former scenario appears to be more plausible since we obtain $\beta_d \approx 0.26$ for all $d$~\cite{jiricek2023critical}  and $\beta_C^{\,(d)}\approx 0.33$ for all $d$ in Figure~\ref{figA6}. 
\section{Connection to Fractal Dimension}
\label{sec:scaleinvariance_3}

Finally, let us comment on the relationship between the power-law exponents $\beta$ and the fractal dimension.
In Refs.~\cite{hopjan2023, hopjan2023scaleinvariant}, it was established that the power-law exponent $\beta$ of the decay of survival probability is related to the fractal dimension $\gamma$, which is defined via the finite-size scaling of the inverse participation ratio (see Refs.~\cite{hopjan2023, hopjan2023scaleinvariant} for details).
The precise relation is given by $\gamma=n\beta$, where $n$ determines the scaling of the typical Heisenberg time with the system size at the critical point, $t_H^{\rm typ} = D^n$ ($n\approx 1$ in the 3D Anderson model and $n\approx 2$ in the 1D Aubry--André model~\cite{hopjan2023}).
If the dynamics of particle imbalance are nearly identical to the dynamics of survival probability, as in the case studied here for the typical initial product states, then one should expect the analogous relation $\gamma_I = n \beta_I$ with $\gamma_I = \gamma$ and $\beta_I = \beta$.
Note that in Ref.~\cite{jiricek2023critical}, the relation $\gamma_I = n \beta_I$ was found even for the initial CDW states.
However, while for the initial CDW states it was observed that $\gamma_I \approx \gamma$ in the 3D Anderson model, it was also observed that $\gamma_I \neq \gamma$ in the 1D Aubry--André model.

Here, we check for both models whether a similar relation also holds true for the equal-time connected density--density correlation functions,
\begin{equation} \label{def_gamma_n_C}
\gamma_C^{\,(d)} = n \beta_C^{\,(d)}.\;
\end{equation}
Numerical results for $\beta_C^{\,(d)}$  are shown in the main panels of Figures \ref{figA5} and~\ref{figA6}, while the numerical results for $\gamma_C^{\,(d)}$ are shown in the corresponding insets.
For the 1D Aubry--André model, the relation in Equation (\ref{def_gamma_n_C}) is valid up to the second decimal digit for all $d$, as can be seen from Figure~\ref{figA6}.
In the case of the 3D Anderson model, the relation in Equation (\ref{def_gamma_n_C}) is valid almost up to the second digit for $d=0$, see Figure~\ref{figA5}a, but it deteriorates when $d$ is increased, see Figure \ref{figA5}b--d.
We checked that a similar agreement is observed also for the transition probabilities. 
 We believe that the deviations at large $d$ are finite size effects since the numerical calculation of the slopes requires two fitting procedures, for which the results get less predictive as we increase $d$ while keeping $L$ the same.
The validity of Equation (\ref{def_gamma_n_C}) may be interpreted as being a consequence of the scale-invariant power-law behavior. 

\begin{adjustwidth}{-\extralength}{0cm}
\reftitle{References}

\PublishersNote{}
\end{adjustwidth}
\end{document}